\documentclass[twocolumn]{aastex61}
\pdfoutput=1 
\usepackage{amsmath,amstext}
\usepackage[T1]{fontenc}
\usepackage{apjfonts} 
\usepackage[figure,figure*]{hypcap}
\usepackage{appendix}

\shorttitle{Feedback in Gas Rich Disks}
\shortauthors{Fisher et al.}

\begin{document}


\title{Testing Feedback Regulated Star Formation in Gas Rich, Turbulent Disk Galaxies}

\author{Deanne~B.~Fisher}
\affiliation{Centre for Astrophysics and Supercomputing, Swinburne
  University of Technology, P.O. Box 218, Hawthorn, VIC 3122,
   Australia}
   
\author{Alberto~D.~Bolatto}
 \affiliation{Laboratory of Millimeter Astronomy, University of Maryland, College Park, MD 29742}
 
\author{Heidi White}
\affiliation{Department of Astronomy \& Astrophysics, University of Toronto, 50 St. George St., Toronto, ON M5S 3H8, Canada}

\author{Karl Glazebrook}
\affiliation{Centre for Astrophysics and Supercomputing, Swinburne
 University of Technology, P.O. Box 218, Hawthorn, VIC 3122,
Australia}
\affiliation{ARC Centre of Excellence for All-sky Astrophysics (CAASTRO)} 
   
\author{Roberto G. Abraham}
\affiliation{Department of Astronomy \& Astrophysics, University of Toronto, 50 St. George St., Toronto, ON M5S 3H8, Canada}
  

\author{Danail Obreschkow}
\affiliation{International Centre for Radio Astronomy Research (ICRAR), M468, University of Western Australia, 35 Stirling Hwy, Crawley, WA 6009, Australia}


   


\begin{abstract}
In this paper we compare the molecular gas depletion times and mid-plane hydrostatic pressure in turbulent, star forming disk galaxies to internal properties of these galaxies. For this analysis we use 17 galaxies from the DYNAMO sample of nearby ($z\sim0.1$) turbulent disks. 
We find a strong correlation, such that galaxies with lower molecular gas depletion time ($t_{dep}$) have higher gas velocity dispersion ($\sigma$). Within the scatter of our data, our observations are consistent with the prediction that $t_{dep}\propto \sigma^{-1}$ made in theories of feedback regulated star formation. We also show a strong, single power-law correlation between mid-plane pressure ($P$) and star formation rate surface density ($\Sigma_{SFR}$), which extends for 6 orders of magnitude in pressure.  Disk galaxies with lower pressure are found to be roughly in agreement with theoretical predictions. However, in galaxies with high pressure we find $P/\Sigma_{SFR}$ values that are significantly larger than theoretical predictions. Our observations could be explained with any of the following: (1) the correlation of $\Sigma_{SFR}-P$ is significantly sub-linear; (2) the momentum injected from star formation feedback ($p_*/m_*$) is not a single, universal value; or (3) alternate sources of pressure support are important in gas rich disk galaxies. Finally using published survey results, we find that our results are consistent with the cosmic evolution of $t_{dep}(z)$ and $\sigma(z)$. Our interpretation of these results is that the cosmic evolution of $t_{dep}$ may be regulated not just by the supply of gas, but also the internal regulation of star formation via feedback.   
\end{abstract}


\section{Introduction}
The majority of star-formation in massive galaxies occurred roughly 10~billion years ago, during the epoch $z\sim 1-3$
\citep{hopkinsbeacom2006,madau2014}. There are significant differences in the gas and star formation properties in those distant galaxies, compared to local universe spirals.  
The molecular gas fraction increases significantly from $\sim0$ to $z\sim 3$ \citep[e.g.][]{tacconi2013,combes2013}, while  molecular gas depletion time,
$t_{dep} \equiv M_{mol}/SFR$, has a only modest decrease from $\sim$2~Gyr at
$z\approx0$ \citep{leroy2012,bigiel2008,rahman2012,saintonge2011b} to
$\sim0.3-0.7$~Gyr at $z\approx2$ \citep{tacconi2013,genzel2015,scoville2016,schinnerer2016,magdis2017}.  

The observed evolution of molecular gas depletion time is considerably
more shallow than predictions from simulations and semi-analytic
models \citep[e.g.][]{dave2011,lagos2015}. These theories assume that the depletion time is mostly regulated by available gas supply, and the cosmic evolution of the dynamical time. Alternatively, the slow evolution has been seen as evidence that local processes may determine gas depletion times \citep[e.g][]{genzel2015}. Indeed, the internal properties of galaxies at higher redshift are observed to be very different. They have super-giant star forming regions \citep{genzel2011,wisnioski2012,guo2012,fisher2017mnras} and elevated gas velocity dispersions \citep[e.g.]{forsterschreiber2009,lehnert2009,swinbank2012,wisnioski2015}. 
 Recent high-resolution observations confirm predictions \citep[e.g.][]{dekel2009clumps,genzel2011} that the properties of these giant star-forming regions, so-called ``clumps'', are directly consistent with being the result of galaxy wide disk instabilities \citep{fisher2017apjl,white2017,dessauges2018}.  We note that alternate theories do exist. For example \cite{inoue2018} argue that clumps are consistent with forming via fragmentation of spiral arms, which are likewise consistent with data. 
 
 The elevated gas velocity dispersions in gas rich galaxies are most commonly interpreted as signatures of strong, galaxy wide turbulence. Some authors have interpreted this turbulence to be driven by star formation feedback from star formation \citep[e.g.][]{lehnert2013,green2010}, however the nature of the mechanism driving this turbulence remains under debate \citep[e.g.][]{krumholz2016}. 

\cite{ostriker2010} present a detailed model for star formation in which the vertical pressure is balanced by pressure supporting mechanisms, such as energy injected from supernovae. In this model turbulence is driven primarily by feedback from supernova explosions. Models in which star formation feedback drives turbulence \citep[also][]{faucher2013} predict that in marginally stable systems the depletion time is inversely dependent on the vertical velocity dispersion ($\sigma_z$). \cite{ostriker2011}  also predicts that the SFR surface density should be directly proportional, or at least nearly proportional \citep{kim2013,shetty2012}, to the mid-plane pressure of the galaxy disk. In essence the pressure from the galaxy is supported by the energy injected from feedback processes associated with star formation. If the depletion time is indeed linked to the internal kinematics this may give an explanation for the slow cosmic evolution of $t_{dep}$, as it would be regulated not just by gas inflow but also by feedback processes. 

\cite{krumholz2012} present an argument in which the turbulence is driven by gravitation alone. They argue that feedback is not necessary to describe the bulk properties of star formation in galaxies. \cite{krumholz2016} argue that galaxies follow a linear correlation such that $SFR \propto \sigma$, albeit with significant scatter, which is predicted in this theory. These models predict a very different dependence of depletion time. In this case, $t_{dep}$ is most affected by the dynamical time of the galaxy \citep[similar to ][]{dave2011,lagos2015}, and either has no dependence on velocity dispersion or a positive dependence. Different models of turbulence driving mechanisms therefore predict very different parameter dependencies with $t_{dep}$. 

Heretofore, these direct predictions remain difficult to test due to a lack of sufficient range in parameters. For example, $\sigma$ is effectively constant across well studied samples of nearby galaxies, like THINGS \citep{leroy2008}. Observations of higher redshift galaxies would provide larger dynamic range in properties like mid-plane pressure, molecular gas depletion time and $\sigma$. However, with present facilities the signal-to-noise of internal properties of galaxies is low, and introduces both a significant amount of scatter to correlations, as well as a selection bias towards larger, brighter targets.

\begin{deluxetable*}{lcccccccc}
\tablewidth{0pt} \tablecaption{Sample Properties}
\tablehead{ \colhead{Galaxy} & \colhead{z} & \colhead{M$_{star}$} & \colhead{M$_{mol}$} & \colhead{Kinematic}  & \colhead{SFR}& \colhead{$R_{1/2}$} &\colhead{Emission} &\colhead{Gas} \\
\colhead{ }& \colhead{ } &\colhead{$10^{10} M_{\odot}$ } & \colhead{$10^9 M_{\odot}$} &\colhead{Source\tablenotemark{a} } & \colhead{$M_{\odot}$~yr$^{-1}$  } & \colhead{kpc} & \colhead{Line\tablenotemark{a}} &\colhead{Source\tablenotemark{a}}
}
\startdata
C14-2 & 0.0562 & 0.56 & 1.83 $\pm$ 0.36 & WiFeS$^3$ & 1.12 $\pm$ 0.17 & 4.0 & H$\alpha$$^3$  & IR SED$^3$ \\
D00-2 & 0.0813 & 2.43 & 5.08 $\pm$ 0.84 & WiFeS$^3$ & 5.14 $\pm$ 0.72 & 3.5 & H$\alpha$$^3$ & IR SED$^3$ \\
G03-2 & 0.12946 & 0.65 & 5.16 $\pm$ 1.18 & WiFeS$^3$ & 4.6 $\pm$ 0.89 & 4.5 &H$\alpha$$^3$  & IR SED$^3$  \\
G10-1 & 0.14372 & 2.75 & 13.45 $\pm$ 2.15 & GMOS$^3$ & 15.7 $\pm$ 1 & 1.2 & H$\alpha$$^3$  & CO(1-0)$^3$ \\
D20-1 & 0.07049 & 2.95 & 5.93 $\pm$ 0.65 & WiFeS$^3$ & 4.7 $\pm$ 0.25 & 3.4 & H$\alpha$$^3$  & CO(1-0)$^3$ \\
G04-1 & 0.12981 & 6.47 & 29.00 $\pm$ 2.10 & GMOS$^2$ & 21.32 $\pm$ 1 & 2.8 & H$\alpha$$^2$ & CO(1-0)$^1$ \\
G20-2 & 0.14113 & 2.16 & 5.22 $\pm$ 0.59 & GMOS$^2$ & 18.24 $\pm$ 0.35 & 2.1 & H$\alpha$$^2$ & CO(1-0)$^3$ \\
D13-5 & 0.07535 & 5.38 & 11.90 $\pm$ 0.36 & GMOS$^2$ & 17.48 $\pm$ 0.45 & 2.0 & H$\alpha$$^2$ & CO(1-0)$^1$ \\
G08-5 & 0.13217 & 1.73 & 7.11 $\pm$ 0.79 & GMOS$^2$ & 10.04 $\pm$ 1 & 1.8 &H$\alpha$$^2$ & CO(1-0)$^3$ \\
D15-3 & 0.06712 & 5.42 & 9.36 $\pm$ 0.18 & WiFeS$^2$ & 8.29 $\pm$ 0.35 & 2.2  & H$\alpha$$^2$ & CO(1-0)$^3$ \\
G14-1 & 0.13233 & 2.23 & 4.94 $\pm$ 0.59 & GMOS$^2$ & 6.9 $\pm$ 0.5 & 1.1 & H$\alpha$$^2$& CO(1-0)$^3$  \\
C13-1 & 0.07876 & 3.58 & 5.91 $\pm$ 0.15 & WiFeS$^2$ & 5.06 $\pm$ 0.5 & 4.2 & H$\alpha$$^2$ & CO(1-0)$^3$ \\
C22-2 & 0.07116 & 2.19 & 4.94 $\pm$ 0.35 & OSIRIS$^4$ & 3.2 $\pm$ 0.28 & 3.4 & Pa~$\alpha$$^4$  & CO(1-0)$^5$ \\
SDSS024921-0756 & 0.153 & 3.02 & 7.30 $\pm$ 0.56 & OSIRIS$^4$ & 10.54 $\pm$ 1.054 & 1.1 & Pa~$\alpha$$^4$ & CO(1-0)$^5$  \\
SDSS212912-0734 & 0.184 & 7.08 & 51.15 $\pm$ 3.77 & OSIRIS$^4$ & 53 $\pm$ 5.3 & 1.3 & Pa~$\alpha$$^4$  & CO(1-0)$^5$  \\
SDSS013527-1039 & 0.127 & 7.08 & 34.89 $\pm$ 1.61 & OSIRIS$^4$ & 25.27 $\pm$ 2.527 & 1.6 & Pa~$\alpha$$^4$ & CO(1-0)$^5$   \\
SDSS033244+0056 & 0.182 & 7.24 & 40.32 $\pm$ 2.39 & OSIRIS$^4$ & 60.5 $\pm$ 6.05 & 1.9 & Pa~$\alpha$$^4$ & CO(1-0)$^5$   \\
\enddata
\tablenotetext{a}{References: $^1$-\cite{fisher2014}, $^2$-\cite{fisher2017apjl}, $^3$-\cite{white2017}, $^4$-\cite{oliva2017}, $^5$- This publication}

\end{deluxetable*}

We use the
DYNAMO sample \citep{green2010,green2014} of rare galaxies located
at $z=0.075\sim0.2$ that have properties very similar to turbulent,
clumpy disk galaxies more commonly found at higher redshift.

Throughout this paper,
we assume a concordance cosmology with \hbox{$H_0$ = 67 km\ $s^{-1}$ Mpc$^{-1}$}, $\Omega_M=0.31$,
and $\Omega_\Lambda=0.69$.

\section{Sample and Data Sources}
\begin{figure}
\begin{center}
\includegraphics[width=0.48\textwidth]{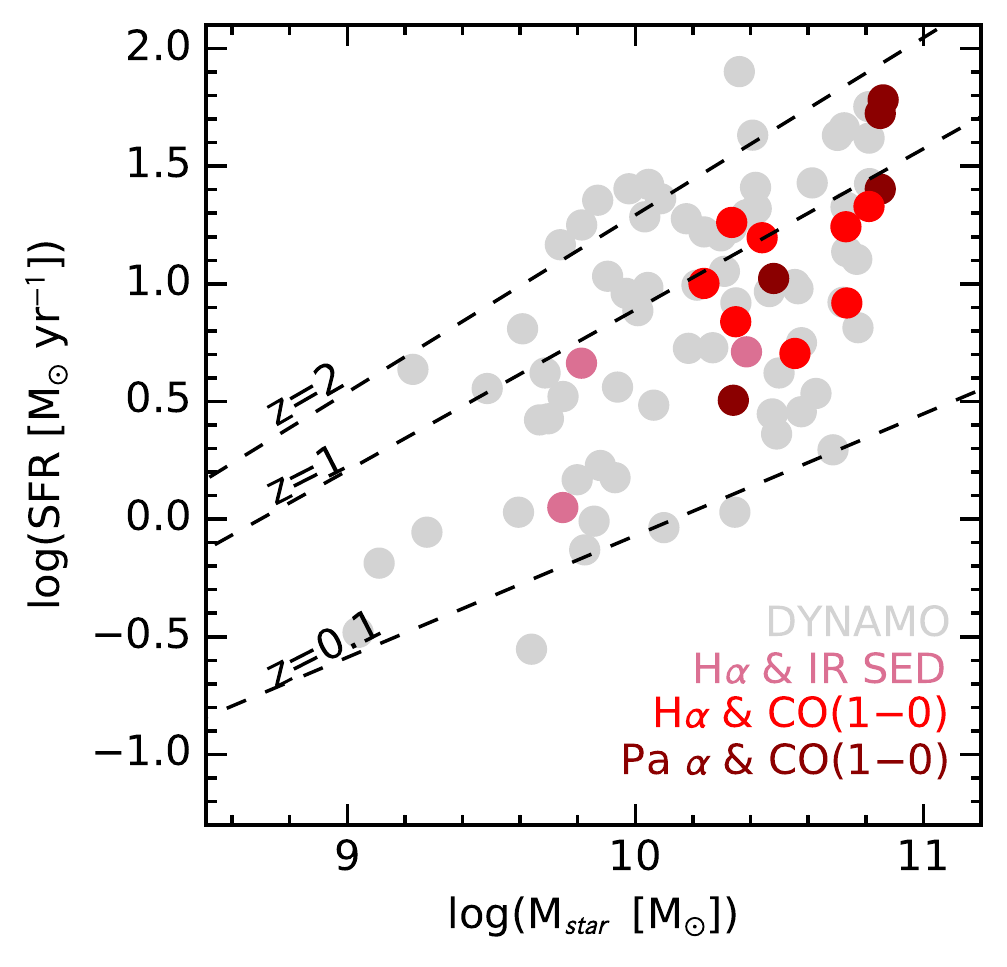} 
\end{center}
\caption{SFR is plotted against stellar mass for galaxies in our sample. The color of data points indicates the data type used to determine the SFR and the gas mass. We also plot as dashed lined the main-sequence of galaxies at three different redshifts, $z=0.1,1,2$. Main-sequence values are taken from \cite{speagle2014}. DYNAMO galaxies span the range in $SFR$-M$_{*}$ from $z>0.1$ to $z\leq 2$. \label{fig:ms} }
\end{figure} 
\subsection{DYNAMO Sample}
The galaxies considered here are a subset of the DYNAMO survey galaxies \cite{green2014}. DYNAMO galaxies are selected from SDSS DR4 based on high H$\alpha$ line flux, while excluding AGNs from the sample. 
In Fig.~\ref{fig:ms} we show that the galaxies in our sample have specific star-formation rates ($SFR/M_{*}$) that range from the low-$z$ main-sequence values to those of $z\approx$1-2. The sample spans $M_{star} = 
0.9-9\times 10^{10}$~M$_{\odot}$, $SFR\sim 1-60$~M$_{\odot}$~yr$^{-1}$ and extinction $A(H\alpha)\sim
0.2-1.7$~mag. Overall, galaxies similar to those in DYNAMO-{\em HST} are extremely rare in the local Universe, with a space density of $\sim 10^{-8}-10^{-7}$~Mpc$^{-3}$.

A number of observational results have been published showing that DYNAMO galaxies are more similar to $z\approx 1.5$ main-sequence galaxies than local Universe ULIRGs (Ultra-Luminous Infrared Galaxies). DYNAMO
galaxies have very high molecular gas fractions, $f_{gas}\approx
0.2-0.8$ \citep{fisher2014,white2017}, where as local Universe galaxies typically have molecular gas fractions of less than 10\%. \cite{white2017} shows that unlike local Universe ULIRGs, DYNAMO galaxies have lower dust temperatures ($T_{dust}\sim 20-30$~K).

An important similarity of DYNAMO galaxies to $z\sim 1-2$ main-sequence galaxies is in the kinematics. DYNAMO galaxies are rotating systems with high gas velocity dispersions, $\sigma\sim
20-100$~km~s$^{-1}$ 
\citep{green2010,green2014,bassett2014,bekiaris2016}. Observations with $\sim$100~pc resolution confirm these high dispersions are not caused by beam-smearing effects \citep{oliva2017}. Moreover, \cite{bassett2014} show that these kinematic signatures are also observed in the stellar kinematics of DYNAMO galaxies, thus indicating it is not likely a gas disk inside of system of stars with different kinematics. Finally, \cite{obreschkow2015} show that the angular momentum of DYNAMO galaxies is low for disks of their stellar mass, but is very similar to what has since been observed in $z\approx 1.5$ main-sequence galaxies \citep{swinbank2017}.   

Using HST H$\alpha$
maps, \cite{fisher2017mnras} show that DYNAMO galaxies are ``clumpy'',
and when DYNAMO images are degraded to match
$z\approx2$ observations they meet quantified definitions of clumpy
galaxies \citep[e.g.][]{guo2015}.  Moreover,  \cite{fisher2017apjl} show that those DYNAMO galaxies identified as ``disks'' meet detailed predictions of \cite{toomre1964} instability theories, whereas control galaxies identified as mergers do not.  Specifically, the correlation between the size of clumps and kinematics of the disk. 

Overall, the properties of DYNAMO galaxies most closely resemble galaxies at $z\approx1-2$. DYNAMO galaxies are therefore excellent laboratories for studying the processes in clumpy, turbulent disks with higher resolution and greater sensitivity. 

In this analysis we only include targets that are identified as "disk galaxies". We use the same criteria as described in \cite{fisher2017mnras} that disk galaxies show both rotating ionized gas in 2-D velocity fields and are well fit by an exponentially decaying surface brightness profile. For surface photometry we use the FR647M continuum images for galaxies imaged with HST and the 1.9~$\mu$m continuum for galaxies observed with OSIRIS. For the galaxy G10-1 we use 500~nm continuum from GMOS for the stellar surface brightness profile, this has 1.2~kpc resolution. Galaxies G03-2, D00-2 and C14-2 only have SDSS imaging to measure the stellar surface brightness profile, which is considerably poorer resolution. However, all three of these galaxies have measured dust temperatures from Herschel data of 20-30~K \citep{white2017}, which strengthens the case that they are disk-like systems rather than major-mergers. These 3 targets are plotted as different color points in all results figures.  As discussed in \cite{fisher2017mnras} galaxies H10-2 and G13-1 do not meet the criteria of disks, and from \cite{oliva2017} galaxy E23-1 does not meet this definition of a disk galaxy.  

\begin{deluxetable*}{lcccccc}
\tablewidth{0pt} \tablecaption{Derived Properties of Sample Galaxies }
\tablehead{ \colhead{Galaxy} & \colhead{$\sigma $ } & \colhead{$V_{flat} $}  & \colhead{$t_{dep} $} & \colhead{$\Sigma_{SFR}$} & \colhead{$P/k_B$} & \colhead{$P_{DE}/k_B$} \\
\colhead{ } & \colhead{ [km s$^{-1}$] } & \colhead{[km s$^{-1}$] } & \colhead{[Gyr]} &\colhead{ [log($M_{\odot}~yr^{-1}~kpc^{-2}$)] } & \colhead{ [log(cm$^{-3}$ K)]  } & \colhead{[log(cm$^{-3}$ K)] }  
}
\startdata
C14-2 & 26 & 159 & 1.63 $\pm$ 0.41 & -2.254 $\pm$ 0.093 & 4.63 $\pm$ 0.42 & 4.96 $\pm$ 0.37 \\
D00-2 & 35 & 61 & 0.99 $\pm$ 0.21 & -1.464 $\pm$ 0.089 & 5.47 $\pm$ 0.28 & 5.80 $\pm$ 0.21  \\
G03-2 & 32 & 189 & 1.12 $\pm$ 0.34 & -1.751 $\pm$ 0.106 & 4.91 $\pm$ 0.41 & 5.27 $\pm$ 0.31   \\
G10-1 & 52 & 117 & 0.86 $\pm$ 0.15 & -0.094 $\pm$ 0.071 & 7.64 $\pm$ 0.24 & 7.96 $\pm$ 0.15   \\
D20-1 & 35 & 134 & 1.26 $\pm$ 0.15 & -1.500 $\pm$ 0.032 & 5.56 $\pm$ 0.21 & 5.89 $\pm$ 0.17   \\
G04-1 & 50 & 269 & 1.36 $\pm$ 0.12 & -0.650 $\pm$ 0.030 & 6.94 $\pm$ 0.16 & 7.26 $\pm$ 0.13 \\
G20-2 & 81 & 166 & 0.29 $\pm$ 0.03 & -0.482 $\pm$ 0.023 & 6.45 $\pm$ 0.18 & 6.82 $\pm$ 0.13 \\
D13-5 & 46 & 192 & 0.68 $\pm$ 0.03 & -0.475 $\pm$ 0.024 & 6.86 $\pm$ 0.14 & 7.18 $\pm$ 0.12  \\
G08-5 & 64 & 243 & 0.71 $\pm$ 0.11 & -0.628 $\pm$ 0.048 & 6.67 $\pm$ 0.19 & 7.04 $\pm$ 0.13 \\
D15-3 & 45 & 240 & 1.13 $\pm$ 0.05 & -0.867 $\pm$ 0.028 & 6.61 $\pm$ 0.13 & 6.94 $\pm$ 0.12 \\
G14-1 & 70 & 136 & 0.72 $\pm$ 0.10 & -0.358 $\pm$ 0.038 & 7.33 $\pm$ 0.19 & 7.68 $\pm$ 0.14  \\
C13-1 & 29 & 223 & 1.17 $\pm$ 0.12 & -1.644 $\pm$ 0.048 & 5.27 $\pm$ 0.25 & 5.60 $\pm$ 0.20 \\
C22-2 & 32 & 164 & 1.54 $\pm$ 0.17 & -1.647 $\pm$ 0.044 & 5.45 $\pm$ 0.25 & 5.78 $\pm$ 0.19 \\
SDSS024921-0756 & 57 & 84 & 0.69 $\pm$ 0.09 & -0.135 $\pm$ 0.061 & 7.58 $\pm$ 0.16 & 7.91 $\pm$ 0.13 \\
SDSS212912-0734 & 53 & 105 & 0.97 $\pm$ 0.12 & 0.411 $\pm$ 0.061 & 8.56 $\pm$ 0.16 & 8.82 $\pm$ 0.13 \\
SDSS013527-1039 & 41 & 232 & 1.38 $\pm$ 0.15 & -0.110 $\pm$ 0.061 & 7.88 $\pm$ 0.16 & 8.14 $\pm$ 0.13 \\
SDSS033244+0056 & 59 & 239 & 0.67 $\pm$ 0.08 & 0.134 $\pm$ 0.061 & 7.81 $\pm$ 0.14 & 8.11 $\pm$ 0.13  \\
\enddata
\tablenotetext{a}{Uncertainty on $\sigma$ is 3-5 km s$^{-1}$ and on $V_{flat}$ is 5-10 km s$^{-1}$. }
\end{deluxetable*}

For our analysis we make observations to measure molecular gas masses, ionized gas maps with $\sim100$~pc resolution, and kinematics of ionized gas. We present gas masses and kinematics on 17 DYNAMO galaxies, and we obtain ionized gas maps for 13 targets. All of these methods are well tested and have been used on the DYNAMO sample in previously published works. Here we summarize the basics of each.

\subsection{Gas Masses}
 We compile CO(1-0) observations from \cite{fisher2014} and \cite{white2017} using PdBI and NOEMA, respectively, with new NOEMA observations. 
 
New NOEMA observations were made during the period May-December 2016, with observing programs S16CK and W16BK (PI:Fisher). We use the same observational method and similar measurement techniques as was previously published in \cite{fisher2014} and \cite{white2017}.  Similar to previous observations \citep{fisher2013,white2017} all observations were made with the {\em WIDEX} correlator. These observations comprise 6 of the targets listed in Table~1. 

Typical integration times were $1\sim 2$ hours in C or D configurations. The correlator was tuned to observe the redshifted CO(1-0) emission line in each target. This yielded an on-sky frequency of 97-107~GHz for targets ranging in redshifts from 0.18 to 0.07. These observations were then calibrated using standard {\em GILDAS} routines in {\em CLIC}, and then cleaned with the {\em MAPPING} pipeline routine  during an on-site visit to IRAM.  Observations were reduced with the default channel width of 20~km~s$^{-1}$, and then binned to 50~km~s$^{-1}$ yielding cubes with typical RMS$\sim2-3$~mJy. Final flux uncertainties for each target are given in Table~3.  

Our targets are unresolved in these PdBI and NOEMA observations. Two of the sources (D~20-1 and C22-2) have particularly elongated beams. However, these do not significantly affect the measurement of the flux. We have overlaid the beam shape and size on SDSS r-band images and checked each source for possible contamination from other sources. The only source with an additional source in the beam area is G20-2, which contains a point source representing less than 1\% of the flux of G20-2 in the FR647M HST continuum image, but is barely detectable in H$\alpha$. It is, therefore, not likely this is changing the CO flux by a significant amount.  

The spectrum for each target is measured in a polygon region containing the galaxy. These spectra are shown in in the Appendix. Fluxes are obtained by binning the data into 50~km~s$^{-1}$ channels, and then integrating the resulting spectrum. The choice in range of channels to integrate over is made by starting at the redshifted frequency of the CO(1-0) line and summing all adjacent channels that are above the noise limit. The total line-widths of our targets are typically 300-400~km~s$^{-1}$. Noise levels for the observations ranged 0.5-1.0~mJy. We also consider that the choice in how measure flux may affect the final value. We therefore also measure flux by fitting a single component Gaussian function to the data, as well as measuring the flux in 20~km~s$^{-1}$ channels. We take the standard-deviation of the 4 different fluxes for each target and add this in quadrature with the noise in the spectrum to estimate the measurement uncertainty of flux. This value is shown in each spectrum in the Appendix. The range of signal-to-noise for new NOEMA observations was $S/N=10-22$. Observational details of CO detections are listed in Table~3.

The CO(1-0) flux is converted to molecular gas mass ($M_{mol}$)
in the usual fashion, in which $M_{mol} = \alpha_{CO} L_{CO}$, where
$L_{CO}$ is the luminosity of CO(1-0), and $\alpha_{CO}$ is the
CO-to-H$_2$ conversion factor, including a 1.36$\times$ correction for
heavier molecules. We adopted the standard value $\alpha_{CO}=4.36$. We have made multiple efforts to determine the most appropriate value of $\alpha_{CO}$. First, based on SDSS spectra DYNAMO galaxies have slightly sub-solar metallicity. Recently, \cite{white2017} study the dust temperature for a set of DYNAMO galaxies, including several in this work. They find that DYNAMO galaxies have low dust temperatures, $T_{dust}\sim 20-30$~K, implying Milky Way like conversion factors \citep[as reviewed in][]{bolatto2013}. Moreover, of the few galaxies that have both dust SED and CO(1-0) observations the estimated gas masses agree to $\sim$25\%. Finally, \cite{fisher2014} uses the formula from \citep{bolatto2013} which estimates $\alpha_{CO}$ via the  baryonic surface density, and found a results consistent with the Milky Way value.

We also include 3 targets from \cite{white2017} for which the gas mass is determined from {\em Herschel} observations. Here we estimate the dust mass by fitting black-body models to the infrared spectral energy distribution, and then converted to molecular gas mass by assuming a constant dust-to-gas ratio, similar to other works \citep[e.g.][]{genzel2015}.

\subsection{Star Formation Rates}
\cite{fisher2017mnras} presents 10 HST H$\alpha$ maps of DYNAMO galaxies, using ramp filters FR716N and
FR782N using the Wide Field Camera on
the Advanced Camera for Surveys (WFC/ACS) on the Hubble Space
Telescope ({\em HST}; PID 12977, PI Damjanov). Integration times were
45~min in the narrow band filter and 15~min with the continuum
filter. All images were reduced using the standard {\em HST} pipeline. We correct the fluxes measured in the image using a an [NII]/H$\alpha$ ratio determined from the SDSS spectrum for each target.  Seven of the the DYNAMO-HST sample galaxies also have CO(1-0) fluxes and are included here. A detailed description of the observations, continuum subtraction and clump measurement is given in \cite{fisher2017mnras}. The typical resolution of DYNAMO-HST observations is 60-150~pc. 

We also observed 5 galaxies with Pa~$\alpha$ emission line observed with Keck OSIRIS \citep{larkin2006}. Our targets were observed with the laser guide star system with exposure times of 4$\times$900~s, achieving resolution of 150-400~pc. We use the OSIRIS data reduction pipeline version 2.3. To flux-calibrate the OSIRIS spectra, we use the telluric stars observed each night (an average of 3 telluric stars per night). Our method corresponds to a first order flux calibration consistent with the 2MASS magnitudes within $\sim$20\%. Detailed descriptions of both observation and reduction of OSIRIS data cubes is given in \cite{bassett2017} and \cite{oliva2017}.

Star formation rates are calculated from the emission line flux using the extinction-corrected H$\alpha$ line luminosity by assuming star
formation rate~[M$_{\odot}$~yr$^{-1}$] = $5.53\times 10^{-42}
L_{H\alpha} $ [~erg~s$^{-1}$] \citep{hao2011}. We calculate the
intrinsic H$\alpha$ extinction using the H$\alpha$ and H$\beta$ line
ratios from SDSS spectrum. For those targets with OSIRIS data we divide the Pa~$\alpha$ luminosity by the intrinsic luminosity ratio, $L_{Pa~\alpha}/L_{H\alpha}=0.128$ \citep{calzetti2001}. Using H$\alpha$ fluxes from \cite{green2017} we find that $SFR_{H\alpha} - SFR_{Pa\alpha}\approx \pm 2.5$~M$_{\odot}$~yr$^{-1}$. 

Three galaxies in our sample have neither HST nor OSIRIS data. For these we use the published H$\alpha$ emission line fluxes measured from AAT/SPIRAL and AAT/WiFES \citep{green2017}.

\subsection{Kinematics}
Data cubes containing intensity, velocity dispersion and
rotation velocity of each galaxy are extracted from the data by fitting Gaussian functions with PSF convolution to emission lines in individual spaxels. These are then fit with kinematic models
by method of least-squares using the GPU based software {\em gbkfit} \citep[see][]{bekiaris2016}. Inclination is taken from photometry. We model the rotation velocity, $v_{rot}$, with the
function\citep{boissier2003}
\begin{equation}
v_{rot}(r) = v_{flat} \left [ 1 - exp(-r/r_{flat}) \right].
\end{equation} 
The software then returns $v_{flat}$ and $r_{flat}$.

The fit to the data cubes also includes an intrinsic component of velocity dispersion, $\sigma$. In the model the velocity dispersion is assumed to be constant across the disk. Fitting the velocity dispersion simultaneously in the model allows for accounting of the beam smearing in the data \citep{davies2011}, as well as inclination. This flat velocity dispersion profile makes a necessary assumption that the galaxy is a disk, and further requires our effort to exclude mergers as described above.

An alternative approach to modeling velocity dispersions is to make a weighted average of the velocity dispersion in the region of the galaxy that also shows a flat rotation curve. \cite{oliva2017} investigates both methods with DYNAMO galaxies. They generate model galaxies designed to match clumpy DYNAMO disks. Then fit these with both {\em gbkfit} as well as straightforward averages. They find that both methods recover similar values for velocity dispersion, with modeling being slightly more stable as it intrinsically accounts for rising dispersion in galaxy centers. On average the models and average methods have a difference of $\sigma_{ave} - \sigma_{model}\lesssim 5$~km~s$^{-1}$, which is consistent with our error bars.  
 
The data sources for both $\sigma$ and $V$ include AAT/WiFES H$\alpha$ observations \citep{green2014}, Keck OSIRIS observations of Pa~$\alpha$ \citep{oliva2017}, and Gemini GMOS observations of H$\alpha$ and H$\beta$ \citep{bassett2014,fisher2017apjl}. For a more detailed description of Gemini observations see \cite{bassett2014}. In those cases in which multiple observations are made on a single target we preference first the GMOS observations, as this data set offers both deep high S/N observations and sub-kpc resolution, then OSIRIS observations due to the high resolution, then AAT observations. The kinematic parameters we use here, derived from the three separate data sets are found to generally agree on the order of the uncertainties, $\sim\pm5-10$km~s$^{-1}$ \citep{bekiaris2016,bassett2014,oliva2017}.

\section{Molecular Gas Depletion Time in Turbulent Disks}
In our sample we find a range of $t_{dep}= 0.3 - 1.6$~Gyr, with the average for our sample at 1~Gyr. This range in $t_{dep}$ was the intended effect of targeting galaxies with a range in SFR/M, as shown in Fig.~\ref{fig:ms}. Our observations reproduce the  relationship\footnote{Throughout this paper we measure parameter correlations using the the maximum likely hood $R$ package {\em Hyperfit} \citep{robotham2015}. } between $t_{dep}$ and the SFR/M$_{*}$ as \cite{saintonge2011b} of $t_{dep}\propto (SFR/M)^{-0.53\pm0.14}$ with Pearson's correlation coefficient of $r=-0.6$. This correlation has also been observed in high redshift galaxies \citep[e.g.][]{tacconi2017}. This continues to motivate our treatment of this DYNAMO sample as well representing the properties of actively star forming galaxies. 

\begin{figure}
\begin{center}
\includegraphics[width=0.5\textwidth]{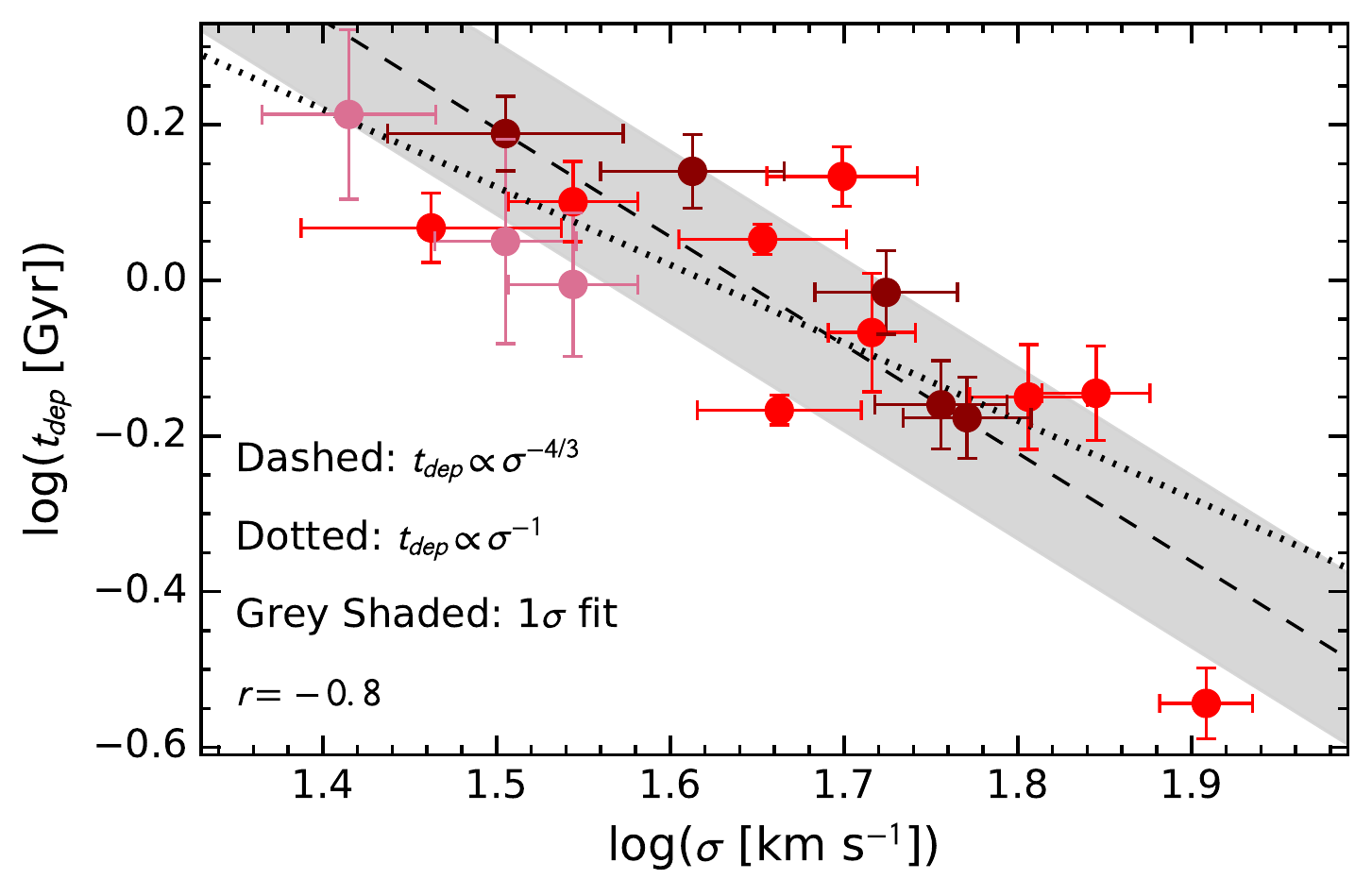} 
\end{center}
\caption{ The relationship between galaxy depletion time, and internal gas velocity dispersion is shown. Symbol colors represent the source of data as described  in Fig.~\ref{fig:ms}. The grey region indicates the 1$\sigma$ scatter around the best-fit relation. The dashed line represents the prediction from multi-free fall, turbulence models \citep{salim2015}. The dotted line indicates the prediction from feedback driven models \citep[e.g.][]{ostriker2010}. Indeed we observe a strong negative correlation between $t_{dep}$ and $\sigma_{gas}$.  \label{fig:sig}  }
\end{figure}

\subsection{Relationship with Gas Velocity Dispersion}
In Fig.~\ref{fig:sig} we compare $t_{dep}$ to velocity dispersion of ionized gas($\sigma$).  For reference, \cite{wisnioski2015} finds that $\sigma_{gas}$ increases from $\sim10-20$~km~s$^{-1}$ in the local Universe to $\sigma>30$~km~s$^{-1}$ at $z>1$. In our sample, all galaxies with $t_{dep}<1$~Gyr have $\sigma>40$~km~s$^{-1}$. 

Considering our entire DYNAMO data set we find a strong, inverse relationship between the molecular gas depletion time and the gas velocity dispersion, with Pearson's correlation coefficient of $r=-0.8$ and $p$-value of $8.7\times10^{-5}$. The best fit relationship for these quantities is 
\begin{equation}
\log(t_{dep}) = -1.39\pm0.23 \times \log(\sigma) + 2.27\pm0.38,
\label{eq:sig}
\end{equation}
where $t_{dep}$ is in Gyr and $\sigma$ is in km~s$^{-1}$. The vertical scatter around this correlation is 0.12~Gyr. We note in our data set this correlation has less scatter, and a stronger correlation coefficient than that of $t_{dep}$ and SFR/M$_{*}$.

To investigate how much the correlation 
depends on the galaxy with the largest velocity dispersion/shortest 
depletion timescale (G20-2), we remove it from the sample and recompute $r$ and $p$.   It is possible that this galaxy is at the least affecting the power-law of the correlation to some degree. We therefore refit the data excluding galaxy G20-2. We still find a strong, inverse correlation for the data set that excludes this target, with Pearson's correlation coefficient $r=-0.78$ and $p$-value of $4.7\times10^{-4}$. The best fit quantities for this subset of the data that does not include $G20-2$ is 
\begin{equation}
\log(t_{dep}) = -1.04\pm0.20 \times \log(\sigma) + 1.71\pm0.33.
\label{eq:sig2}
\end{equation}

To be clear the exclusion of G20-2 is a completely {\em ad hoc} treatment. We do not observe any special features of this target that lead us to believe it is any more peculiar than the other DYNAMO galaxies. \cite{fisher2017mnras} presents both the H$\alpha$ map and 600~nm continuum surface brightness profile of all DYNAMO galaxies, and G20-2 does not have significant asymmetries, aside from the presence of clumps. Moreover, there are no detectable companion galaxies in the HST images.  We do note that G20-2 has a very prominent ring with radius $\sim$1~kpc. This may be driving a higher star formation efficiency or larger $\sigma$ when compared to other DYNAMO galaxies. However, G04-1 and D13-5 both have rings as well, albeit less prominent than that of G20-2.  We note that in Fig.~\ref{fig:p_sigSFR} of this work we will find that G20-2 is not an out-lier.

As discussed in \cite{robotham2015} the maximum likelihood fitting technique provides a robust treatment of uncertainties. Nonetheless, we also consider ordinary least squares (OLS) bisector fits to the data. We do this to ensure that our fitting method is not somehow biasing the result. Also, OLS techniques have been in use for a much longer time \citep{isobe1990}, allowing for standard comparison with previous work. We find that a fit to all our data recovers an OLS bisector of $t_{dep}\propto \sigma^{-1.30\pm0.18}$ and the fit to the data set in which G20-2 is omitted recovers an OLS bisector of  $t_{dep}\propto \sigma^{-1.03\pm0.19}$. These different fitting methods, therefore, yield results that are within uncertainties of each other.

The error bars in Fig.~\ref{fig:sig} are representative of the measurement uncertainties propagated through to the physical quantities. In the case of depletion time, however, it is likely that the systematic uncertainty is somewhat larger. The systematic uncertainty on $t_{dep}$ could be as high as a factor of 2 for any single object \citep[eg.][]{bolatto2013}. To determine the impact of this on the robustness of the correlation in Fig.~\ref{fig:sig} we carry out a simple bootstrap experiment. We randomly modify the $log(t_{dep})$ values  to scatter around each point within a Gaussian distribution with $\sigma_{gauss}= 0.15$~dex. We ran 1000 iterations, determining the correlation coefficient and $p$-value of each realization. We find that the median correlation coefficient is $r=-0.63$ with a standard deviation of 0.11, and the median $p=0.006$ with standard deviation of 0.04. We rerun this with more pessimistic assumption of a flat distribution with width of $\pm$0.15~dex, and find similar median correlation coefficient of $r=-0.6$ and p-value still indicating a strong correlation, with median $p=0.01$. Removing G20-2 from the fit reduces the robustness of the correlation, but with $r=-0.53$ and $p=0.03$ the data still represents a high probability of correlation. Therefore, the correlation we observe in Fig.~\ref{fig:sig} appears to be robust against the systematic uncertainties on the depletion time.

As we state above, exclusion of G20-2 in the fitted sample only marginally affects the robustness of the fit, measured either by Pearson's $r$ or the $p$-value. Both sample choices satisfy definitions of "strong correlation" based on these statistical metrics. However, inclusion (or exclusion) of G20-2 does have a significant impact on the power-law in the correlation between $t_{dep}$ and $\sigma$. Future samples that contain more galaxies with $t_{dep}\lesssim 0.5$~Gyr would be helpful to further constrain the exact value of the power-law.  With our current sample there appears to be enough evidence to support a statistically significant inverse correlation between $t_{dep}$ and $\sigma$, with power-law ranging between $\sigma^{-1}$ and $\sigma^{-1.4}$. We will therefore consider predictions within this range consistent with our data set.  

We note that the systematic observational effect of increased velocity dispersion on $\alpha_{CO}$ should result in the opposite trend. If the velocity dispersion of gas is increased by components other than internal cloud properties, then the effect is a systematic increase in observed CO luminosity \citep[see discussion in][]{bolatto2013}. This would increase the observed $t_{dep}$ in galaxies with larger $\sigma$. Our observations of the opposite trend in Fig.~\ref{fig:sig}, then implies this correlation is likely physical, and may even be steeper than our observations if considers 
variations in $\alpha_{CO}$. 

Before we compare our observational results to predictions from theory we point out an important assumption. Our measurement of velocity dispersion relies on ionized gas velocity dispersion maps, whereas theoretical predictions refer total gas or molecular gas velocity dispersions. Ionized gas velocity dispersions includes broadening due to thermal broadening. For a gas of 10$^4$~K this broadening amounts to $\sigma_{br}\approx 10$~km~s$^{-1}$, and is added in quadrature with the line-width due to the motion of gas, such that the observed velocity dispersion is $\sigma_{obs}^2\approx \sigma_{gas}^2 + \sigma_{br}^2$. For galaxy samples with lower measured velocity dispersions, such as in dwarf galaxies \citep[as shown in][]{Moiseev2015} this can be a significant contribution, however, our observed velocity dispersions range from 26 to 81~km~s$^{-1}$. Thermal broadening will at most contribute 1-2~km~s$^{-1}$. 

\cite{levy2018} find in nearby spiral galaxies that ionized gas gives systematically lower rotation velocities than molecular gas. They argue that this is because in low-z spiral molecular gas is in a thin disk, where the ionized gas traces a thicker, more turbulent component. However, it is not clear if the same result holds for galaxies, like DYNAMO and z=1-2 main-sequence galaxies, that have significantly higher surface densities of gas and star formation.  \cite{white2017} argues that for systems in dynamical equilibrium, which have large gas mass surface densities, the bulk of the gas will naturally have higher scale heights, which they show are well represented by the velocity dispersions measured with ionized gas in DYNAMO galaxies. Ultimately this field would significantly benefit from direct comparisons of ionized and molecular gas kinematics in gas rich, turbulent disks, such studies are at present sparse. Recent work by \cite{ubler2018} compares ionized gas and molecular gas kinematics in a single star forming disk galaxy at z=1.4. They find that the kinematics, both rotation and velocity dispersion, for the two tracers agree well, within 1-5~km~s$^{-1}$. More work on this comparison would certainly be welcome, nonetheless, the evidence thus far seems to suggest that at high SFR and high $\sigma_{ion}$ the ionized gas is a good tracer of the gas kinematics. We therefore make the same assumption that as other studies \citep[e.g.][]{forsterschreiber2009,lehnert2009,green2010,genzel2011,wisnioski2015,krumholz2016} that for our sample $\sigma_{ion}\approx\sigma_{gas}$.

Theories of the ISM that incorporate feedback from star formation predict coupling of $t_{dep}$ and $\sigma$ that is similar to our observations \citep{ostriker2010,shetty2012,faucher2013,krumholz2017}. In all of these theoretical derivations it is shown that in the limit of marginal stability the turbulent velocity has a linear relationship with the star formation efficiency. If we assume that in our targets the ionized gas velocity dispersion is mostly driven by turbulence, then these models predict $t_{dep}\propto \sigma^{-1}$. This is represented as a dotted line Fig.~\ref{fig:sig}, and is consistent with our data, assuming an arbitrary scale factor. In subsequent a section we will return to the topic of the scale-factor in the $\sigma-t_{dep}$ correlation in light of results presented there.  


\cite{salim2015} derive a star formation law using a multi-freefall prescription of the gas \citep[see also][]{federrath2012}. They predict that star formation efficiencies will depend on both the probability density distribution and the sonic Mach number of the turbulence. In the limit that the virial parameter is not significantly variable, the Mach number is directly proportional to the velocity dispersion. These models then predict $t_{dep}\propto \sigma^{-4/3}$, which is shown as a dashed line in Fig.~\ref{fig:sig}. The prediction based on the multi-freefall model is almost exactly the same as our best fit relation to our entire data set, $t_{dep}\propto\sigma^{-1.39}$.  

In the comparison of both the theories of feedback regulates star formation and the multi-free fall model one must consider the variation the  free-fall time, $t_{ff}$. A relationship between $\sigma$ and $t_{dep}$ in star formation theory can be taken from \cite{shetty2012} Equation 20, which states that $\sigma \propto (t_{ff}/t_{dep}) (p_*/m_*)$. Therefore, our interpretation of Fig.~\ref{fig:sig} assumes that $t_{ff}$ varies less than $t_{dep}$ and $\sigma$. Similarly, with the results of \cite{salim2015}, the simple relationship $t_{dep} \propto \sigma^{-3/4}$ can only be derived by assuming that one can neglect variation in $t_{ff}$. 

For the sample of galaxies in Fig.~\ref{eq:sig} the we expect that there significant variation in the galaxy averaged free fall time is unlikely. We expect that $t_{ff}\propto 1/\sqrt{\rho}$ \citep{krumholz2005}; where $\rho$ is the volume density. We can estimate the variation in $\rho$ using the parameters $\Sigma_{gas}/h_{z}$, where $h_z$ is the gas scale-height. For galaxies with longer $t_{dep}\sim 1-2$~Gyr the typical surface density in our sample is 10-50~M$_{\odot}$~pc$^{-2}$. We surmise that these lower $\sigma$ galaxies likely have a disk thickness that is similar to the Milky Way, of order $\sim$100~pc. For a clumpy galaxy with lower $t_{dep}$ and higher $\sigma$ the gas surface density is typically 100-1000~M$_{\odot}$~pc$^{-2}$. For these we assume a disk thickness that is similar to $z\sim1$ edge-on galaxies of 500-1000~pc \citep{elmegreen2005b}. \cite{bassett2014} also presents a discussion of two DYNAMO disk thicknesses based on kinematics of stars and gas; they conclude that $h_z$ is in the range 400-1000~pc. Even though the galaxies in the bottom left portion of Fig.~2 have higher surface densities, they are also thicker, and therefore the value of $1/\sqrt{\Sigma_{gas}/h_z}$  will not change much across the sample in Fig.~\ref{fig:sig}.\ This is consistent with the results of \cite{krumholz2012}, who find that the distributions of free-fall times of "high-z disks" and "z=0 spirals" overlap. Moreover, any variance in $t_{ff}$ over this small dynamic range is likely to be significantly affected by stochastic scatter.  However, if we extended our sample either to very lower surface density disks or perhaps very extreme star-bursting disks at $z\sim4$ this assumption that $t_{ff}$ does not significantly vary may not be as valid. We note that there is significant uncertainty on the scale-height of the molecular disk in gas-rich disk galaxies, and more work needs to be undertaken to study this important quantity. We also note that these are galaxy-averaged quantities. It is very likely that measurements at the scale of individual molecular clouds have increased scatter, and perhaps a different parameter dependency due to the more significant variation of the free-fall time.

\begin{figure}
\begin{center}
\includegraphics[width=0.5\textwidth]{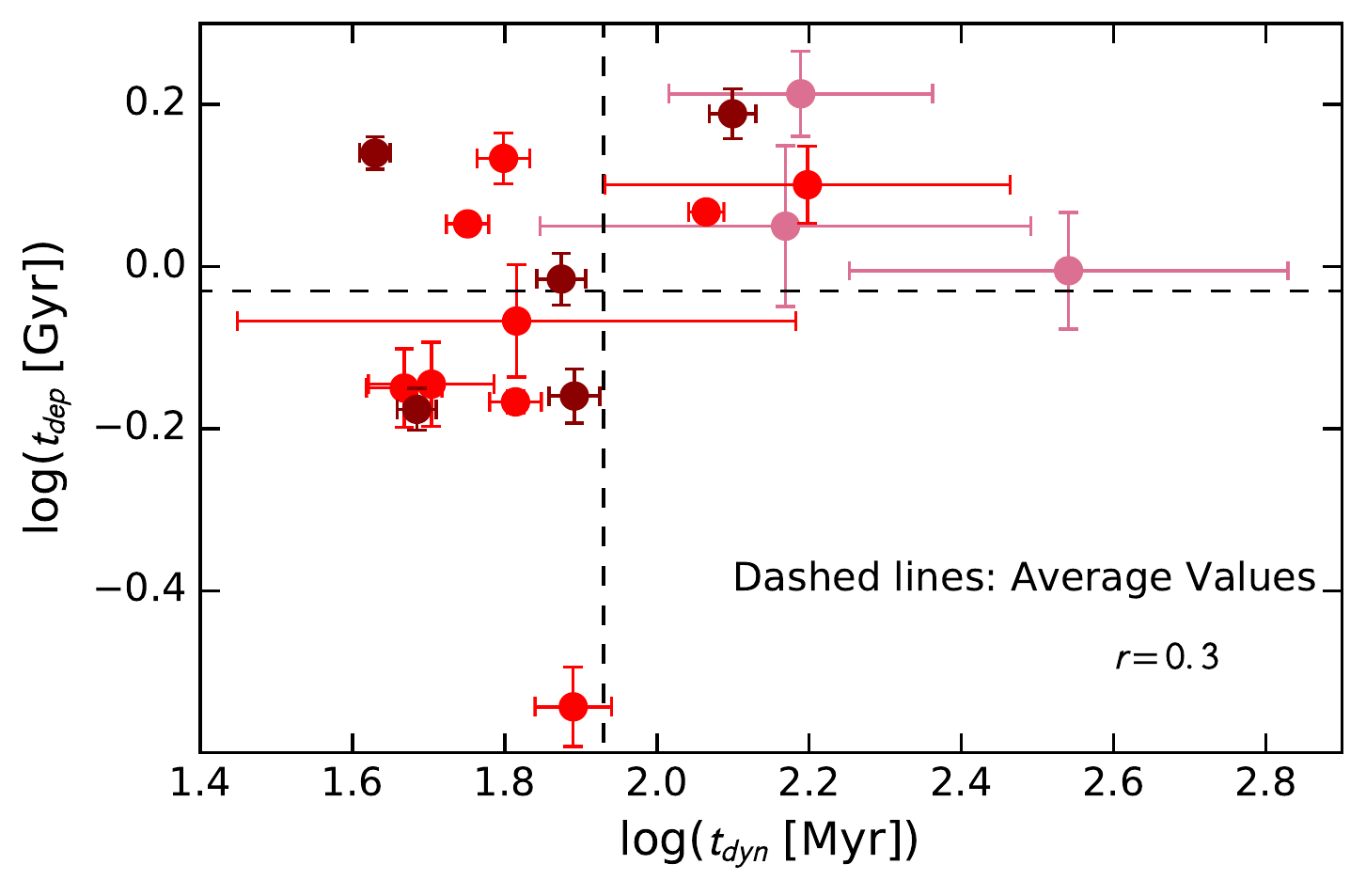} 
\end{center}
\caption{ The relationship between galaxy depletion time and dynamical time. Symbol colors represent the source of data as described  in Fig.~\ref{fig:ms}. The dashed lines indicate the averages for each quantity. We observe a very weak correlation ($r=-0.30$). The best fit is $t_{dep}\propto t_{dyn}^{0.44}$, with considerable scatter. \label{fig:tdyn}  }
\end{figure}

\subsection{Dynamical Time}
A number of theories and semi-analytic models suggest that the gas depletion time should be directly connected to the dynamical time, $t_{dyn}= 2\pi R_{disk}/V_{flat}$, of the galaxy \citep[e.g.][]{dave2011,krumholz2012}. These models are motivated by the well-known relationship $\Sigma_{SFR}\propto \Sigma_{gas}/t_{dyn}$ observed in nearby galaxies \citep{kennicutt98}. \cite{krumholz2012} argues that in the "Toomre regime", in which galaxies have high gas fractions and show marginal stability, the local star forming region is not able to decouple from the ambient gas in the galaxy. From this concept they then derive a positive, linear correlation between $t_{dyn}$ and $t_{dep}$.

In Fig.~\ref{fig:tdyn} we show the relationship between depletion time and the galaxy dynamical time for galaxies in our sample.  The depletion time does have some dependency on $t_{dyn}$ in that galaxies with  $t_{dep}\lesssim 1$~Gyr always have low $t_{dyn}$. Effort to fit a correlation returns a very weak, high scatter relationship, with Pearson's correlation coefficient of $r=0.3$ and $p$-value for this data set is 0.2. The best fit relationship for our data set is 
\begin{equation}
\log(t_{dep}) = 0.44\pm0.24 \log(t_{dyn}) -0.44\pm0.23.
\end{equation}
This is significantly more shallow than the prediction for gravity driven turbulence \citep{krumholz2012}. We find no correlation at all between $t_{dep}$ and the product of $Q\times t_{dyn}$, which is also predicted in the gravity-only theory. We find that using the same bootstrap method as used to analyze Fig.~2, to account for systematic uncertainty in $t_{dep}$, results in a median correlation coefficient of r=0.25 with standard deviation of 0.16, the median p-value is 0.3. These values indicate a very low probability of a correlation between $t_{dep}$ and $t_{dyn}$ in our data set.

\section{Effect of Extreme Pressure on Star Formation in Turbulent Disks} 

Motivated by the qualitative success of feedback regulated star formation models in describing the relationship between $t_{dep}$ and $\sigma$, we now consider a fundamental prediction of those same models, the relationship between $\Sigma_{SFR}$ and mid-plane hydrostatic pressure, $P$. Theories describing the formation of compact star clusters in very high pressures have been in development for some time \citep{elmegreen1989,elmegreen1997}, and observations of gas-rich, star forming disk galaxies suggests pressures can become very high compared to low-z spirals \citep{swinbank2011}. 

It is proposed in a number of models that star formation in disk galaxies is a self-regulating process, in which the pressure of the system is balanced by the feedback processes associated with star formation \citep[e.g.][]{ostriker2011, kim2013,shetty2012}. 
The semi-analytic models in \cite{ostriker2011} predict a linear relationship between pressure and $\Sigma_{SFR}$, and the simulations of \cite{kim2013} find a nearly linear relationship. 

\subsection{Estimating the Total Mid-Plane Pressure}
In this work we use the following formula from \cite{elmegreen1989} to estimate the mid-plane hydrostatic pressure within our galaxies, 

\begin{equation}
P = \frac{\pi}{2} G \Sigma_{g}\left [ \Sigma_g + \left(\frac{\sigma}{\sigma_*}\right)\Sigma_* \right]. \label{eq:mpp}
\end{equation}

$\Sigma_{gas}$ and $\Sigma_{*}$ represent the gas and stellar mass surface densities, and similarly $\sigma$ and $\sigma_*$ represent velocity dispersions of the gas and stars. 

In the Appendix we outline our method to measure or estimate each of these parameters described in Equation~5, as well as discuss in detail the impact of the uncertainty in each parameter on the pressure. We briefly summarize our main sources of uncertainty here. We also note that Equation~5 was originally developed to describe sub-galactic measurements of the ISM. Our use of it here requires the assumption that the ensemble averages of the galaxy are reflective of local values, though systematic biases may exist. 

\begin{figure}
\begin{center}
\includegraphics[width=0.45\textwidth]{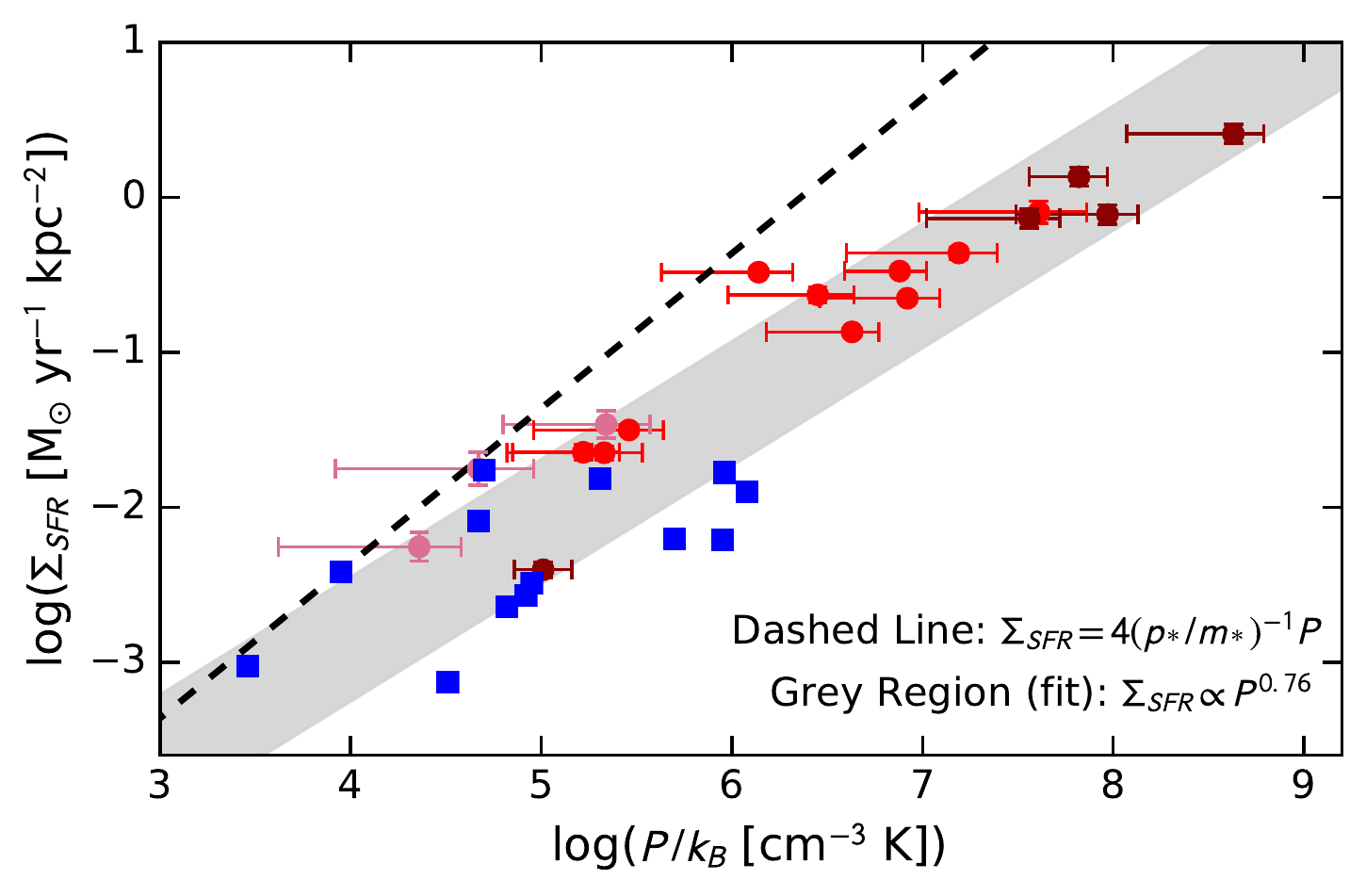} 
\end{center}
\caption{ The above figure compares SFR surface density to mid-plane hydrostatic pressure. The DYNAMO galaxies are labeled according to data source as described in Fig.~\ref{fig:ms}. In this figure we also include measurements from \cite{leroy2008} using THINGS data. The dashed line represents the theoretical prediction.  We find a strong, sub-linear relationship spanning 4-5 orders of magnitude inn both $\Sigma_{SFR}$ and pressure. 
 \label{fig:p_sigSFR}  }
\end{figure}

The largest source of uncertainty in pressure comes from our use of unresolved CO flux measurements. To calculate $\Sigma_{gas}$ we assume that the size of the ionized gas disk is equivalent to the size of the molecular gas disk. This assumption is consistent with observations of high-z galaxies \citep{tacconi2013,hodge2015,bolatto2015,dessauges2015} with an uncertainty of 20-50\%.  However, we also consider the possibility that the molecular disk is as large as our unresolved CO measurement. This reduces the pressure by a factor of $\sim$4 in most targets, and is reflected in the error-bars of Fig.~\ref{fig:p_sigSFR}. We also consider the uncertainty introduced from different assumptions of the atomic hydrogen surface density. We find this only has a significant impact on the 2 lowest pressure systems, and is likewise reflected in the error bars. For more information on these uncertainties see the Appendix.

To increase sampling at low pressures, we combine our sample set here with the observations of the THINGS sample \citep{walter2008}, using derived measurements from \cite{leroy2008} to calculate $\Sigma_{SFR}$ and $P$. The combined data set spans nearly 6 orders-of-magnitude in mid-plane pressure and 4 orders-of-magnitude in $\Sigma_{SFR}$. We note there are important differences between the THINGS and DYNAMO measurements of pressure. The THINGS sample has measured values of both molecular and atomic gas mass surface density, where the DYNAMO sample only has observations of molecular gas mass and atomic gas mass surface density is adopted. This is likely important at lower pressure where the ratio of molecular-to-atomic gas mass surface density is lower. Alternatively, THINGS galaxies do not have measurements of ionized gas velocity dispersions.  \cite{leroy2008} adopts  $\sigma\approx 11$~km~s$^{-1}$ as being consistent with their measurements of atomic gas. These considerations are discussed in more detail in the Appendix.  We have five DYNAMO galaxies that overlap the range in derived pressure and $\Sigma_{SFR}$ with those of the THINGS galaxies. We find the two samples to have similar values of $\Sigma_{SFR}/P$.

For DYNAMO galaxies we find very high values of $P/k$ compared to local spirals like the Milky Way, reaching as much as $10^5$ times higher than the pressure in the Milky Way. This is similar to the pressure of the $z\sim2$ galaxy observed in \cite{swinbank2011}. DYNAMO galaxies have both significantly higher gas masses and smaller sizes compared to Milky Way-like spirals. These differences lead to greater surface densities, which then  create very high pressures that we observe.  In almost all galaxies the ``stellar term'' of the pressure, $(\sigma/\sigma_*) \Sigma_*$ dominates over the ``gas term'', $\Sigma_g$. This is notable as the gas fractions of DYNAMO galaxies are very high, $f_{gas}\sim 20 -60\%$. Even in this sample of gas-rich galaxies we find that on average $P_{star}/P_{gas}\approx 2.3\pm1.1$. We note that this is a prediction of feedback-regulated star formation, that even in gas rich systems the stars contribute a very important component to the pressure \citep{ostriker2011}. We will return to this subject in our discussion of dynamical equilibrium pressure.

\subsection{Balance of Star Formation and Pressure in Gas Rich Disks}

In  Fig.~\ref{fig:p_sigSFR} we compare the mid-plane hydrostatic pressure $P$, as described in Equation~5 with the star formation rate surface density. We find that a single correlation describes this data well, over 6 orders-of-magnitude in pressure. The best fit relation is 
\begin{equation}
\log(\Sigma_{SFR}) = (0.76\pm 0.06) \log(P/k_{B}) - 5.89\pm0.35.
\end{equation}
Where $\Sigma_{SFR}$ is in units of M$_{\odot}$~yr$^{-1}$~kpc$^{-2}$, and $P/k_{B}$ is in units of cm$^{-3}~K$. The 1~$\sigma$ vertical scatter around this relationship is represented as the shaded grey region in Fig.~\ref{fig:p_sigSFR}. 

It is predicted that $\Sigma_{SFR} = 4(p_*/m_*)^{-1} P_{tot}^N$, where the power-law index $N$ is expected to be close to unity \citep[e.g.][]{shetty2012,kim2013}. Our correlation in Fig.~\ref{fig:p_sigSFR} is then qualitatively consistent with the prediction of a strong, positive relationship between these two quantities. 

The power-law slope we observe in Fig.~\ref{fig:p_sigSFR} is significantly below unity, which is different than the predictions described above (shown as the dashed line in the figure). In simulations by \cite{kim2013} the measured power-law is slightly steeper than linear. In a subsequent section of this paper we discuss three possible options to explain this discrepancy: (1) That the relationship between pressure and $\Sigma_{SFR}$ is truly non-linear; (2) that the scale-factor $p_*/m_*$ is non-universal and increases for gas-rich, high pressure disks; and/or  (3) that alternate mechanisms, such as mass-transport within a disk, also provide pressure support.

It is important to note that both systematic and observational uncertainties in the calculation of $P/k$ and $\Sigma_{SFR}$ can affect these results at the factor of a few level. For example, to calculate mass surface density we estimate the size of the disk as $R_{disk}=2\ R_{1/2}$. However, since pressure scales as mass surface density squared, adopting a different characteristic radius of the galaxy would alter pressure more than $\Sigma_{SFR}$. The median value for these low pressure systems is $P/\Sigma_{SFR}\approx 9\times10^3$~km~s$^{-1}$, which is only a factor of a few higher than the theoretical prediction, and as shown in Fig.~\ref{fig:p_sigSFR} the predicted value falls within the scatter of low pressure systems. Considering this we therefore conclude that for disks with lower values of pressure ($P/k<10^6$~cm$^{-3}$~K) the data are within uncertainty of the predictions from the models.  This is to say that our data suggests that local Universe spirals are consistent with theoretical predictions of $\Sigma_{SFR}-P$ from feedback regulated star formation models. This is similar to the results of \cite{herreracamus2017} who find that KINGFISH galaxies, with $P/k\sim 10^3.5-10^4.5$ are consistent with predictions from \cite{kim2013}.  

High pressure systems, $P/k>10^6$~cm$^{-3}$~K, however, do not appear to be reconcilable with even generous estimations of the uncertainties. If we consider only those systems with $P/k>10^6$~cm$^{-3}$~K, we find an average value of $<P/\Sigma_{SFR}>\approx 4.4\times 10^4$~km~s$^{-1}$. This is more than an order-of-magnitude larger than the theoretical prediction, as shown  in Fig.~\ref{fig:p_sigSFR}.

In light of these high values of $P/\Sigma_{SFR}$ in DYNAMO galaxies, we now consider the zero-point in Fig.~\ref{fig:sig}. We can compare the scale-factor in Fig.~\ref{fig:sig} to that of Fig.~\ref{fig:p_sigSFR} by solving each theoretical prediction for $p_*/m_*$.  \cite{shetty2012} predict that $\sigma \approx 0.366 (t_{ff}/t_{dep}) (p_*/m_*)$, where $t_{ff}$ is the free-fall time. Note this is adopted from Equation 20 of \cite{shetty2012}.  The estimate using $t_{dep}$ and $\sigma$ does not require all of the assumptions that go into measurement of galaxy pressure (described in detail in the Appendix), most notably galaxy sizes are not used in this case. However, this derivation depends on an assumption of the free-fall time, which introduces a source of uncertainty. Nonetheless, we can think of this method of deriving the scale-factor as a semi-independent check. The free-fall time of a cloud will depend on the inverse square root of the local gas density \citep{krumholz2012}. For a gas-rich, turbulent disk \cite{krumholz2012} suggests values of $t_{ff}\approx 1-10$~Myr. Using this range we derive values of $p_*/m_*\approx 10^4 - 10^6$~km~s$^{-1}$ for DYNAMO galaxies, which is consistent with our estimate from $P/\Sigma_{SFR}$. We find that for individual galaxies there is a high-scatter correlation consistent correlation that finds on average the values of $p_*/m_*$ derived from the two methods are in agreement.

We caution that it is not necessarily the case that  $P/\Sigma_{SFR}$ is a unique, robust tracer of the true physical balance of these quantities over the entire range of galaxies. That is to say, systematic uncertainties could affect this correlation at some level. Most notably, the CO-to-H$_2$ conversion factor could be a function of pressure \citep{narayanan2011}.  

\subsection{Dynamical Equilibrium Pressure}

We also consider the ``dynamical equilibrium pressure'' as described in \cite{kim2011}, 
\begin{equation}
P_{DE}\approx \frac{\pi G \Sigma_g^2}{2} + \Sigma_g (2 G \rho_{SD})^{1/2} \sigma 
\label{eq:pde}
\end{equation}
where $\rho_{SD}$ is the total mid-plane density, including dark matter and stars, as defined in \cite{ostriker2011}. The total mid-plane density $\rho_{SD} = \Sigma_{*}/h_z + (V_{flat}/R_{disk})^2/(4\pi G)$. The quantity $h_z$ is the disk thickness; we describe how we calculate it Appendix. As described before we make the simplifying assumption that in galaxies with high gas velocity dispersion, like the DYNAMO sample, the velocity dispersion measured from ionised gas is a good representation of $\sigma$ for the total gas. The dynamical equilibrium pressure is an alternate description of the pressure, under the assumption that the system has evolved to its equilibrium state in which pressure balances the feedback mechanism. 

An interesting feature of representing pressure as done in Equation~\ref{eq:pde} is that this directly ties the result in Fig.~\ref{fig:sig} to the effect of increasing pressure. We note again that this is under the assumption that $\sigma_z \approx \sigma_{los}$.  If vertical dynamical equilibrium in a disk is satisfied then \cite{ostriker2010} predicts that $P\propto \Sigma \sqrt{\rho_{SD}}$, and also as discussed above $\Sigma_{SFR}\propto P_{DE}$. Therefore in order for both a linear (or nearly linear) relationship between pressure and also $t_{dep} \propto 1/\sigma$ to both hold the stellar gravity term ($\sigma \sqrt{2G\rho_{SD}}$) from Equation~\ref{eq:pde} must dominate over the gas gravity term ($\pi G \Sigma/2$). Therefore a direct prediction of the feedback-regulated theory of star formation in star-bursting disk galaxies, which we can test with our data, is that $\pi G \Sigma/2 < \sigma\sqrt{2G\rho_{SD}}$. We find that for all galaxies in the DYNAMO sample the stellar gravity term dominates over the gas term. We find that on average the ratio of star-to-gas terms is $\sim$7. The lowest values of star-to-gas terms are in systems with, as one would expect, larger gas fractions, $M_{mol}/M_{dyn}\sim 30-80$\%, however the ratio remains above one reaching down to as low as $\sim$3 for galaxies in our sample.

\section{Discussion of Results}
\subsection{Inverse Correlation of $t_{dep}$ and $\sigma$}
In Fig.~\ref{fig:sig} we show a strong correlation ($r\approx-0.8$ and p-value=$10^{-4}$) between the molecular gas depletion time and velocity dispersion of the ionized gas, such that $t_{dep}\propto \sigma^{-1.39\pm0.23}$. We have tested this correlation against the systematic uncertainties introduced from the scaling of both ionized gas emission lines to star formation rates as well as converting CO(1-0) line flux to molecular gas mass. We find the correlation in our sample to be robust. Indeed we find that $t_{dep}$ correlates more strongly  with $\sigma$ than any other parameter we compared it to (e.g. SFR/M, $\Sigma_{SFR}$, $t_{dyn}$, $\Sigma_{star}$).
 
The inverse correlation between $t_{dep}$ and $\sigma$ seems inconsistent with predictions in which the turbulence is driven only by gravity, that do not incorporate more complex treatments of turbulence \citep[e.g.][]{federrath2012}.
\cite{krumholz2012}, in their Equation~18, determine that for a system in which the turbulence is driven exclusively by gravity the $t_{dep}\propto Q t_{dyn}$, where $Q$ is Toomre's stability parameter and $t_{dyn}$ is the dynamical time of the galaxy.  The depletion time of marginally stable disks, such as our sample, is therefore predicted to mostly be driven by the dynamical time, however our data set does not support a strong correlation between $t_{dyn}$ and $t_{dep}$.  We note the caveat that though there is very little evidence to support a galaxy averaged relationship between $t_{dep}$ and $t_{dyn}$, \cite{krumholz2012} suggest that a local relationship may be stronger (where $t_{dep}$ and $t_{dyn}$ are measured in radial bins). This is not possible to measure with our current data set; resolved observations of CO in turbulent disks would be very helpful to this end.

Theories in which turbulence is driven by feedback \citep[e.g.][]{ostriker2011,shetty2012,faucher2013}, however, predict an inverse relationship $t_{dep}\propto \sigma^{-1}$. As we show in Fig.~\ref{fig:sig}, this power-law slope is consistent with our observations.

Using a multi-freefall time-scale prescription for the gas also agree with our results \citep{federrath2012,salim2015}. The critical parameters in these models include the probability density function, and sonic Mach number of the gas. These models are not necessarily inconsistent with feedback driven models, in that feedback could still drive the compressive form of turbulence, which then produces a different distribution of freefall times.

\begin{figure}
\includegraphics[width=0.45\textwidth]{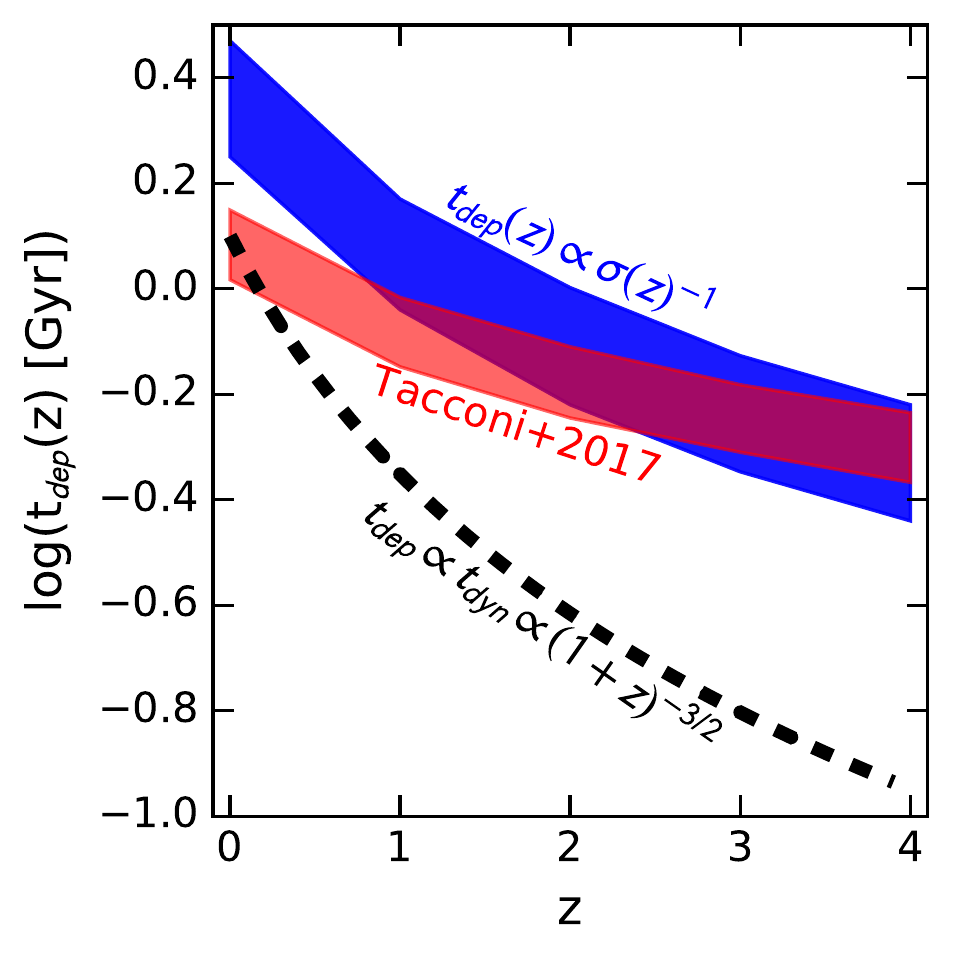}
\caption{In the above figure we consider the very simple toy-model in which the redshift evolution of the depletion time is a function of the cosmic evolution of the velocity dispersion. The blue shaded region represents depletion times as determined from a $\sigma^{-1}$ dependency, similar to that found in Fig.~\ref{fig:sig}. The values of $\sigma(z)$ are taken from \cite{wisnioski2015}. The dashed line represents the expected evolution if $t_{dep}$ is determined only by the dynamical time. The red shaded region represents the cosmic evolution of depletion time from the composite data set of \cite{tacconi2017}. This simple model is a good match to observations at $z>1$. \label{fig:tdepz}}
\end{figure}

\subsubsection{Possible Implications for Cosmic Evolution of $t_{dep}$}
The results we find in this paper may shed some light on the observed shallow evolution of depletion time with redshift \citep{genzel2015,scoville2016,schinnerer2016,tacconi2017}. Theories that drive gas depletion time via the cosmic evolution of the dynamical time predict an order of magnitude decrease in $t_{dep}$. However, recent observations of high-redshift galaxies find that depletion times drop by a factor of 2-5 from $z=0$ to $z=4$.  As we have discussed above in the theory of feedback regulated star formation a lower gas depletion time is natural in systems that have both high pressure and high internal velocity dispersion.

In Fig.~\ref{fig:tdepz} we compare the observed cosmic evolution of molecular gas depletion time to the prediction based on a simple model using a $t_{dep}\propto \sigma^{-1}$ dependency set to match our observations in Fig.~\ref{fig:sig}. \cite{wisnioski2015} presents the gas velocity dispersion, from emission lines, over a range in redshift $z\sim 0-3.5$. We use the average cosmic evolution of $\sigma(z)$, from \cite{wisnioski2015}, to determine $t_{dep}(\sigma)$ as a function of redshift.  The observed cosmic evolution of $t_{dep}$ is from the empirical fit to the main-sequence evolution of the composite data set of \cite{tacconi2017}. The predicted values of $t_{dep}$ agree with observation for $z>1$. This implies first that our results likely hold on high-$z$ galaxies, at least in a bulk sense. Moreover, this may imply that star formation efficiencies at high-redshift are not only regulated by the availability of gas, but also by the feedback within the disk.  

At $z\sim0$ our simple model over-predicts the data by a factor of 2-3. This is not at all surprising. Many of the assumptions that go into our analysis are not valid for low-$z$ spirals. (We remind the reader that DYNAMO galaxies are atypical galaxies for the low-z Universe.) First, unlike our galaxies the velocity dispersion for typical low-z spirals is quite low ($\sigma < 20$~km~s$^{-1}$). As we discuss above ionized gas measurements are more significantly affected by thermal broadening at low dispersion, and ionized gas in more typical, low SFR low-z spirals may overestimate the true gas velocity dispersion. Also, at $z\sim0$ the gas almost certainty becomes far more dominated by the atomic component than in $z>1$ galaxies \citep{obreschkow2009}. 

In Fig.~\ref{fig:tdepz} we also show the expected evolution of the depletion time if it were exclusively a function of the dynamical time, $t_{dep}\propto t_{dyn} \propto (1+z)^{-3/2}$ \citep[see arguments in][]{dave2011,krumholz2016}. This predicts significantly more evolution than in observations. Similarly, \cite{lagos2015} predict a very steep evolution to the depletion time. We interpret this to imply that internal processes, such as the regulation of star formation via feedback, are therefore a dominant factor in determining the evolution of the depletion time over the past $\sim$10 billion years.

\subsection{Implications of $\Sigma_{SFR}-P$ correlation}

We find that in our data set $\Sigma_{SFR}$ and the mid-plane pressure of galaxies have a very tight correlation across almost 6 orders-of-magnitude in pressure using two different formulations of mid-plane pressure (Fig.~\ref{fig:p_sigSFR}). These correlations are found to be sub-linear with $\Sigma_{SFR}\propto P^{0.77}$. Our measurements are therefore not as steep as theories that predict $\Sigma_{SFR} \propto P$ \citep{ostriker2010,kim2011}. 

There appears to be three possibilities to reconcile our observations with the theoretical predictions: (1) The relationship between SFR surface density and pressure is sub-linear; (2) the feedback momentum injected into the ISM per unit new stars formed ($p_*/m_*$) changes as a function of the local ISM properties; and/or (3) alternate mechanisms could drive turbulence and provide support against gravitational pressure. 

From our data alone we cannot uniquely distinguish between these scenarios. We will consider these options below. We note to the reader that there is also the possibility that each of these options are contributing to the offsets we observe. 

\subsubsection{Is $\Sigma_{SFR}$ versus $P$ truly non-linear?}
\cite{benincasa2016} simulate feedback regulated star formation, and qualitatively report a sub-linear relationship between $\Sigma_{SFR}$ and pressure. They argue that feedback affects the scale-height of the disk non-linearly, which affects the pressure and gives rise to this sub-linearity. \cite{benincasa2016} do not estimate an actual value to the power-law. So we cannot quantitatively determine if this effect matches our data. If so then one could assume that the normalization in our Equation~6 can be used to estimate the feedback momentum,  $p_*/m_*\approx 2700$~km~s$^{-1}$. This is similar to the commonly adopted value. Quantitative analysis of the $\Sigma_{SFR}-P$ relationship in 3-D simulations would therefore be informative. 

Observational affects could possibly contribute to a non-linearity as well between $\Sigma_{SFR}$ and $P$. \cite{narayanan2012} argue from simulations that the CO-to-H$_2$ conversion factor may be lower in regions of higher molecular gas surface density. We note over the range of normal spirals empirical studies of the CO-to-H$_2$ conversion factor do not find significant variation with gas mass surface density \citep{sandstrom2013}. \cite{bolatto2013} argues that the baryonic surface density could decrease $\alpha_{CO}$, but this would only be at the $\sim$50\% level. It would not fully reconcile our most extreme observations in Figs.~\ref{fig:p_sigSFR} with the theoretical prediction. Moreover, global studies of $z=1-2$ main-sequence galaxies find that the standard Milky Way conversion factor is consistent with dust mass estimates \citep{genzel2015,tacconi2017}. \cite{white2017} find similar results with DYNAMO galaxies. 

\subsubsection{Does $p_*/m_*$ vary from local spirals to high pressure turbulent disks?}
In theoretical predictions the scale-factor in the relationship between pressure and SFR surface density is directly proportional to the momentum feedback per stellar mass, $p_*/m_*$. In the case that feedback is generated by supernova this quantity, $p_*/m_*$, represents the momentum injected into the ISM from supernova per unit stellar mass of new stars formed. It is therefore a critical parameter in models of feedback regulated star formation.  Most theories of feedback regulated star formation use an adopted or calculated value of $p_*/m_*\sim 3000$~km~s$^{-1}$ \citep[e.g.][]{ostriker2011,shetty2012,faucher2013,kim2017}. The dashed line in Fig.~\ref{fig:p_sigSFR} is set to represent the theoretical prediction. 

The non-linearity in Fig.~\ref{fig:p_sigSFR} could be driven by changing values of $p_*/m_*$ across the range of pressures. If this is the case, our data would be consistent with local spirals, such as THINGS galaxies, having values of $p_*/m_*$ that are roughly consistent with theoretical predictions, but higher pressure systems have significantly higher values of momentum injection.   

We also derive similarly high values of $p_*/m_*$ in DYNAMO galaxies from the $t_{dep}-\sigma$ correlation in Fig.~\ref{fig:sig}, which at face value  provides a semi-independent line of evidence that $p_*/m_*$ is changing. \cite{krumholz2016,krumholz2017} has likewise noted that feedback-only models of the ISM have trouble reproducing the large velocity dispersions observed in $z\sim 2$ turbulent disk galaxies. 

The universality of $p_*/m_*$ is currently under some debate. Some recent simulations suggest that clustering of supernovae could increase the momentum input per star formation by factors of $\sim$10$\times$ \citep{gentry2017}. Conversely, \cite{kim2017} find that the injected momentum per mass of star formation would not be significantly higher in  environments with higher number density of supernovae. More recent simulations from \cite{gentry2018arXiv} that model feedback in a 3D, magnetized medium argue that previous results may be due to numerical effects. 

We note that in the highest pressure systems $p_*/m_* = 4 P/\Sigma_{SFR}\approx 10^5$~km~s$^{-1}$, which is higher than even the most extreme predictions \citep{gentry2017}. This may indicate that varying $p_*/m_*$ alone is not able account for the discrepancy with our data. Resolved observations of $\Sigma_{SFR}$ and $P$ in galaxies like DYNAMO would be useful to further understand the range of values of $P/\Sigma_{SFR}$.

\subsubsection{Are alternate sources of pressure support important in gas rich disks?}
The models we consider above assume that feedback is primarily driven by supernova. Physical models that include higher amounts or radiation feedback, for example larger rates of momentum injection due to the inclusion of radiation pressure, photoionization and winds \citep[e.g.][]{hopkins2011model,murray2011,hopkins2014} are able to generate larger velocity dispersions than those that assume feedback is dominated by supernova, such as \cite{ostriker2011}. This too remains under debate;  high-resolution, detailed simulations of molecular clouds \citep{kim2018rp} suggest that the contribution that radiation feedback makes to $p_*/m_*$ would be quite small ($10^1-10^2$~km~s$^{-1}$) compared to supernovae. Moreover, radiation feedback is found to be even less important in high surface density molecular clouds, which the DYNAMO galaxies would likely contain. 

Recently, \cite{krumholz2017} present a model in which pressure support comes from both mass transport and feedback. Similar to \cite{krumholz2016}, they predict a linear relationship between $\sigma$ and $SFR$ that matches data, including DYNAMO galaxies, with significant scatter. It is possible that in galaxies with higher gas mass surface densities, and hence higher pressures, mass transport plays a more important role. It seems plausible that those galaxies with larger gas fractions would have both more available gas and likely experiencing more active accretion, which could drive more turbulence. However, it is not clear how the relationship between $t_{dep}$ and $\sigma$ would in general be affected by such sources of pressure support. In the specific case that energy lost due to turbulent dissipation is equal to energy injected for supernovae \cite{krumholz2017} find, similar to \cite{ostriker2010}, that $t_{dep}\propto \sigma^{-1}$, which is consistent with our main result in Fig.~\ref{fig:sig}. It is not clear what values of $P/\Sigma_{SFR}$ would exist in this mixed model, nor in those including other forms of feedback, e.g. radiation pressure.

\section{Summary}
Overall our results show qualitative agreement with a number of predictions in feedback regulated star formation models \citep[e.g.][]{ostriker2011}. These include: (1) an inverse correlation between molecular gas depletion time and gas velocity dispersion; (2) a strong, positive correlation between SFR surface density and hydrostatic mid-plane pressure (as well as dynamical equilibrium pressure); and (3) the contribution of stars to the pressure dominates over the gas, even in very gas rich ($f_{gas}>50$\%) galaxies. 

We, however, find significant differences in the quantitative details of both the $t_{dep}-\sigma$ correlation and the $\Sigma_{SFR}$ versus pressure correlations. From our data alone we cannot determine if these correlations imply that (a) the true correlation between $\Sigma_{SFR}$ and pressure is non-linear or (b) the momentum injected into the ISM by star formation feedback ($p_*/m_*$) varies from low values in local spirals to very efficient values in high-$z$ turbulent disks. Moreover, higher spatial resolution observations of molecular gas would help reduce the uncertainty on pressure. There is evidence from simulations that one or possibly both of these options may be contributing to the discrepancies between our observations and theory.

We have also shown that the predictions of feedback regulated star formation, if modified to scale similar to DYNAMO galaxies, is able to account for the cosmic evolution of molecular gas depletion time. Going forward comprehensive studies of both kinematics and gas mass will be useful for determining how relationships like that in Fig.~\ref{fig:sig} hold in high-redshift galaxies. For example, at a given redshift does one see both an increase in $\sigma$ and decrease in $t_{dep}$ in the same way as galaxies extend above the main-sequence. We cannot test this with DYNAMO. Of course high-quality data that can robustly control for galaxy morphology, etc., is presently difficult to obtain. Also we note in closing that more exotic possibilities such as a top-heavy IMF in high $\Sigma_{SFR}$ environment \citep[e.g.][]{nanayakkara2017} could reduce SFR and thus flatten out the redshift evolution of $t_{dep}$; more work to investigate this possibility could be informative.   Finally, as stated above, maps of gas in turbulent disk galaxies will be crucial to determine how pressure may be impacting the properties of massive star forming clusters \citep[as described in][]{elmegreen1997}.

\acknowledgments We are thankful to Cinthya Herrera for help in reducing NOEMA data. We are very grateful to Eve Ostriker, James Wadsley and Christoph Federrath for helpful discussions while preparing this manuscript. DBF is thankful to Sarah Busch for technical help. DBF, KG, and SS acknowledge support from Australian Research Council (ARC) Discovery Program (DP) grant DP130101460. DBF acknowledges support from ARC  Future  Fellowship FT170100376. ADB acknowledges partial support form AST1412419. Some of the data presented herein were obtained at the W. M. Keck Observatory, which is operated as a scientific partnership among the California Institute of Technology, the University of California and the National Aeronautics and Space Administration. The Observatory was made possible by the generous financial support of the W. M. Keck Foundation.  This work is based on observations carried out with the IRAM
Plateau de Bure Interferometer. IRAM is supported by INSU/CNRS
(France), MPG (Germany) and IGN (Spain).

\bibliographystyle{yahapj}

\begin{thebibliography}{}
\providecommand\natexlab[1]{#1}
\providecommand\JournalTitle[1]{#1}

\bibitem[{{Bassett} {et~al.}(2014){Bassett}, {Glazebrook}, {Fisher}, {Green},
  {Wisnioski}, {Obreschkow}, {Cooper}, {Abraham}, {Damjanov}, \&
  {McGregor}}]{bassett2014}
{Bassett}, R., {Glazebrook}, K., {Fisher}, D.~B., {et~al.} 2014,
  \href{http://dx.doi.org/10.1093/mnras/stu1029}{\JournalTitle{\mnras}, 442,
  3206}

\bibitem[{{Bassett} {et~al.}(2017){Bassett}, {Glazebrook}, {Fisher},
  {Wisnioski}, {Damjanov}, {Abraham}, {Obreschkow}, {Green}, {da Cunha}, \&
  {McGregor}}]{bassett2017}
---. 2017,
  \href{http://dx.doi.org/10.1093/mnras/stw2983}{\JournalTitle{\mnras}, 467,
  239}

\bibitem[{{Bekiaris} {et~al.}(2016){Bekiaris}, {Glazebrook}, {Fluke}, \&
  {Abraham}}]{bekiaris2016}
{Bekiaris}, G., {Glazebrook}, K., {Fluke}, C.~J., \& {Abraham}, R. 2016,
  \href{http://dx.doi.org/10.1093/mnras/stv2292}{\JournalTitle{\mnras}, 455,
  754}

\bibitem[{{Benincasa} {et~al.}(2016){Benincasa}, {Wadsley}, {Couchman}, \&
  {Keller}}]{benincasa2016}
{Benincasa}, S.~M., {Wadsley}, J., {Couchman}, H.~M.~P., \& {Keller}, B.~W.
  2016, \href{http://dx.doi.org/10.1093/mnras/stw1741}{\JournalTitle{\mnras},
  462, 3053}

\bibitem[{{Bigiel} \& {Blitz}(2012)}]{bigiel2012}
{Bigiel}, F., \& {Blitz}, L. 2012,
  \href{http://dx.doi.org/10.1088/0004-637X/756/2/183}{\JournalTitle{\apj},
  756, 183}

\bibitem[{{Bigiel} {et~al.}(2008){Bigiel}, {Leroy}, {Walter}, {Brinks}, {de
  Blok}, {Madore}, \& {Thornley}}]{bigiel2008}
{Bigiel}, F., {Leroy}, A., {Walter}, F., {et~al.}
  \href{http://dx.doi.org/10.1088/0004-6256/136/6/2846}{2008, 136, 2846}

\bibitem[{{Blitz} \& {Rosolowsky}(2006)}]{blitzrosolowsky2006}
{Blitz}, L., \& {Rosolowsky}, E. \href{http://dx.doi.org/10.1086/505417}{2006,
  650, 933}

\bibitem[{{Boissier} {et~al.}(2003){Boissier}, {Prantzos}, {Boselli}, \&
  {Gavazzi}}]{boissier2003}
{Boissier}, S., {Prantzos}, N., {Boselli}, A., \& {Gavazzi}, G. 2003,
  \href{http://dx.doi.org/10.1111/j.1365-2966.2003.07170.x}{\JournalTitle{\mnras},
  346, 1215}

\bibitem[{{Bolatto} {et~al.}(2013){Bolatto}, {Wolfire}, \&
  {Leroy}}]{bolatto2013}
{Bolatto}, A.~D., {Wolfire}, M., \& {Leroy}, A.~K. 2013,
  \href{http://dx.doi.org/10.1146/annurev-astro-082812-140944}{\JournalTitle{\araa},
  51, 207}

\bibitem[{{Bolatto} {et~al.}(2015){Bolatto}, {Warren}, {Leroy}, {Tacconi},
  {Bouch{\'e}}, {F{\"o}rster Schreiber}, {Genzel}, {Cooper}, {Fisher},
  {Combes}, {Garc{\'\i}a-Burillo}, {Burkert}, {Bournaud}, {Weiss}, {Saintonge},
  {Wuyts}, \& {Sternberg}}]{bolatto2015}
{Bolatto}, A.~D., {Warren}, S.~R., {Leroy}, A.~K., {et~al.} 2015,
  \href{http://dx.doi.org/10.1088/0004-637X/809/2/175}{\JournalTitle{\apj},
  809, 175}

\bibitem[{{Calzetti}(2001)}]{calzetti2001}
{Calzetti}, D. 2001,
  \href{http://dx.doi.org/10.1086/324269}{\JournalTitle{\pasp}, 113, 1449}

\bibitem[{{Catinella} \& {Cortese}(2015)}]{catinella2015}
{Catinella}, B., \& {Cortese}, L. 2015,
  \href{http://dx.doi.org/10.1093/mnras/stu2241}{\JournalTitle{\mnras}, 446,
  3526}

\bibitem[{{Combes} {et~al.}(2013){Combes}, {Garc{\'{\i}}a-Burillo}, {Braine},
  {Schinnerer}, {Walter}, \& {Colina}}]{combes2013}
{Combes}, F., {Garc{\'{\i}}a-Burillo}, S., {Braine}, J., {et~al.} 2013,
  \href{http://dx.doi.org/10.1051/0004-6361/201220392}{\JournalTitle{\aap},
  550, A41}

\bibitem[{{Dav{\'e}} {et~al.}(2011){Dav{\'e}}, {Finlator}, \&
  {Oppenheimer}}]{dave2011}
{Dav{\'e}}, R., {Finlator}, K., \& {Oppenheimer}, B.~D. 2011,
  \href{http://dx.doi.org/10.1111/j.1365-2966.2011.19132.x}{\JournalTitle{\mnras},
  416, 1354}

\bibitem[{{Davies} {et~al.}(2011){Davies}, {F{\"o}rster Schreiber}, {Cresci},
  {Genzel}, {Bouch{\'e}}, {Burkert}, {Buschkamp}, {Genel}, {Hicks}, {Kurk},
  {Lutz}, {Newman}, {Shapiro}, {Sternberg}, {Tacconi}, \& {Wuyts}}]{davies2011}
{Davies}, R., {F{\"o}rster Schreiber}, N.~M., {Cresci}, G., {et~al.} 2011,
  \href{http://dx.doi.org/10.1088/0004-637X/741/2/69}{\JournalTitle{\apj}, 741,
  69}

\bibitem[{{Dekel} {et~al.}(2009){Dekel}, {Sari}, \&
  {Ceverino}}]{dekel2009clumps}
{Dekel}, A., {Sari}, R., \& {Ceverino}, D. 2009,
  \href{http://dx.doi.org/10.1088/0004-637X/703/1/785}{\JournalTitle{\apj},
  703, 785}

\bibitem[{{Dessauges-Zavadsky} \& {Adamo}(2018)}]{dessauges2018}
{Dessauges-Zavadsky}, M., \& {Adamo}, A. 2018,
  \href{http://dx.doi.org/10.1093/mnrasl/sly112}{\JournalTitle{\mnras}, 479,
  L118}

\bibitem[{{Dessauges-Zavadsky} {et~al.}(2015){Dessauges-Zavadsky}, {Zamojski},
  {Schaerer}, {Combes}, {Egami}, {Swinbank}, {Richard}, {Sklias}, {Rawle},
  {Rex}, {Kneib}, {Boone}, \& {Blain}}]{dessauges2015}
{Dessauges-Zavadsky}, M., {Zamojski}, M., {Schaerer}, D., {et~al.} 2015,
  \href{http://dx.doi.org/10.1051/0004-6361/201424661}{\JournalTitle{\aap},
  577, A50}

\bibitem[{{Elmegreen}(1989)}]{elmegreen1989}
{Elmegreen}, B.~G. 1989,
  \href{http://dx.doi.org/10.1086/167192}{\JournalTitle{\apj}, 338, 178}

\bibitem[{{Elmegreen} \& {Efremov}(1997)}]{elmegreen1997}
{Elmegreen}, B.~G., \& {Efremov}, Y.~N. 1997,
  \href{http://dx.doi.org/10.1086/303966}{\JournalTitle{\apj}, 480, 235}

\bibitem[{{Elmegreen} {et~al.}(2005){Elmegreen}, {Elmegreen}, {Rubin}, \&
  {Schaffer}}]{elmegreen2005b}
{Elmegreen}, D.~M., {Elmegreen}, B.~G., {Rubin}, D.~S., \& {Schaffer}, M.~A.
  2005, \href{http://dx.doi.org/10.1086/432502}{\JournalTitle{\apj}, 631, 85}

\bibitem[{{Faucher-Gigu{\`e}re} {et~al.}(2013){Faucher-Gigu{\`e}re},
  {Quataert}, \& {Hopkins}}]{faucher2013}
{Faucher-Gigu{\`e}re}, C.-A., {Quataert}, E., \& {Hopkins}, P.~F. 2013,
  \href{http://dx.doi.org/10.1093/mnras/stt866}{\JournalTitle{\mnras}, 433,
  1970}

\bibitem[{{Federrath} \& {Klessen}(2012)}]{federrath2012}
{Federrath}, C., \& {Klessen}, R.~S. 2012,
  \href{http://dx.doi.org/10.1088/0004-637X/761/2/156}{\JournalTitle{\apj},
  761, 156}

\bibitem[{{Fisher} {et~al.}(2013){Fisher}, {Bolatto}, {Drory}, {Combes},
  {Blitz}, \& {Wong}}]{fisher2013}
{Fisher}, D.~B., {Bolatto}, A., {Drory}, N., {et~al.}
  \href{http://dx.doi.org/10.1088/0004-637X/764/2/174}{2013, 764, 174}

\bibitem[{{Fisher} {et~al.}(2014){Fisher}, {Glazebrook}, {Bolatto},
  {Obreschkow}, {Mentuch Cooper}, {Wisnioski}, {Bassett}, {Abraham},
  {Damjanov}, {Green}, \& {McGregor}}]{fisher2014}
{Fisher}, D.~B., {Glazebrook}, K., {Bolatto}, A., {et~al.} 2014,
  \href{http://dx.doi.org/10.1088/2041-8205/790/2/L30}{\JournalTitle{\apjl},
  790, L30}

\bibitem[{{Fisher} {et~al.}(2017{\natexlab{a}}){Fisher}, {Glazebrook},
  {Abraham}, {Damjanov}, {White}, {Obreschkow}, {Basset}, {Bekiaris},
  {Wisnioski}, {Green}, \& {Bolatto}}]{fisher2017apjl}
{Fisher}, D.~B., {Glazebrook}, K., {Abraham}, R.~G., {et~al.}
  2017{\natexlab{a}},
  \href{http://dx.doi.org/10.3847/2041-8213/aa6478}{\JournalTitle{\apjl}, 839,
  L5}

\bibitem[{{Fisher} {et~al.}(2017{\natexlab{b}}){Fisher}, {Glazebrook},
  {Damjanov}, {Abraham}, {Obreschkow}, {Wisnioski}, {Bassett}, {Green}, \&
  {McGregor}}]{fisher2017mnras}
{Fisher}, D.~B., {Glazebrook}, K., {Damjanov}, I., {et~al.} 2017{\natexlab{b}},
  \href{http://dx.doi.org/10.1093/mnras/stw2281}{\JournalTitle{\mnras}, 464,
  491}

\bibitem[{{F{\"o}rster Schreiber} {et~al.}(2009){F{\"o}rster Schreiber},
  {Genzel}, {Bouch{\'e}}, {Cresci}, {Davies}, {Buschkamp}, {Shapiro},
  {Tacconi}, {Hicks}, {Genel}, {Shapley}, {Erb}, {Steidel}, {Lutz},
  {Eisenhauer}, {Gillessen}, {Sternberg}, {Renzini}, {Cimatti}, {Daddi},
  {Kurk}, {Lilly}, {Kong}, {Lehnert}, {Nesvadba}, {Verma}, {McCracken},
  {Arimoto}, {Mignoli}, \& {Onodera}}]{forsterschreiber2009}
{F{\"o}rster Schreiber}, N.~M., {Genzel}, R., {Bouch{\'e}}, N., {et~al.} 2009,
  \href{http://dx.doi.org/10.1088/0004-637X/706/2/1364}{\JournalTitle{\apj},
  706, 1364}

\bibitem[{{Gentry} {et~al.}(2017){Gentry}, {Krumholz}, {Dekel}, \&
  {Madau}}]{gentry2017}
{Gentry}, E.~S., {Krumholz}, M.~R., {Dekel}, A., \& {Madau}, P. 2017,
  \href{http://dx.doi.org/10.1093/mnras/stw2746}{\JournalTitle{\mnras}, 465,
  2471}

\bibitem[{{Gentry} {et~al.}(2018){Gentry}, {Krumholz}, {Madau}, \&
  {Lupi}}]{gentry2018arXiv}
{Gentry}, E.~S., {Krumholz}, M.~R., {Madau}, P., \& {Lupi}, A. 2018,
  \JournalTitle{ArXiv e-prints},
  \href{http://arxiv.org/abs/1802.06860}{{\sffamily arXiv:1802.06860}}

\bibitem[{{Genzel} {et~al.}(2011){Genzel}, {Newman}, {Jones}, {F{\"o}rster
  Schreiber}, {Shapiro}, {Genel}, {Lilly}, {Renzini}, {Tacconi}, {Bouch{\'e}},
  {Burkert}, {Cresci}, {Buschkamp}, {Carollo}, {Ceverino}, {Davies}, {Dekel},
  {Eisenhauer}, {Hicks}, {Kurk}, {Lutz}, {Mancini}, {Naab}, {Peng},
  {Sternberg}, {Vergani}, \& {Zamorani}}]{genzel2011}
{Genzel}, R., {Newman}, S., {Jones}, T., {et~al.} 2011,
  \href{http://dx.doi.org/10.1088/0004-637X/733/2/101}{\JournalTitle{\apj},
  733, 101}

\bibitem[{{Genzel} {et~al.}(2015){Genzel}, {Tacconi}, {Lutz}, {Saintonge},
  {Berta}, {Magnelli}, {Combes}, {Garc{\'{\i}}a-Burillo}, {Neri}, {Bolatto},
  {Contini}, {Lilly}, {Boissier}, {Boone}, {Bouch{\'e}}, {Bournaud}, {Burkert},
  {Carollo}, {Colina}, {Cooper}, {Cox}, {Feruglio}, {F{\"o}rster Schreiber},
  {Freundlich}, {Gracia-Carpio}, {Juneau}, {Kovac}, {Lippa}, {Naab}, {Salome},
  {Renzini}, {Sternberg}, {Walter}, {Weiner}, {Weiss}, \& {Wuyts}}]{genzel2015}
{Genzel}, R., {Tacconi}, L.~J., {Lutz}, D., {et~al.} 2015,
  \href{http://dx.doi.org/10.1088/0004-637X/800/1/20}{\JournalTitle{\apj}, 800,
  20}

\bibitem[{{Green} {et~al.}(2017){Green}, {Glazebrook}, {Gilbank}, {McGregor},
  {Damjanov}, {Abraham}, \& {Sharp}}]{green2017}
{Green}, A.~W., {Glazebrook}, K., {Gilbank}, D.~G., {et~al.} 2017,
  \href{http://dx.doi.org/10.1093/mnras/stx1119}{\JournalTitle{\mnras}, 470,
  639}

\bibitem[{{Green} {et~al.}(2010){Green}, {Glazebrook}, {McGregor}, {Abraham},
  {Poole}, {Damjanov}, {McCarthy}, {Colless}, \& {Sharp}}]{green2010}
{Green}, A.~W., {Glazebrook}, K., {McGregor}, P.~J., {et~al.} 2010,
  \href{http://dx.doi.org/10.1038/nature09452}{\JournalTitle{\nat}, 467, 684}

\bibitem[{{Green} {et~al.}(2014){Green}, {Glazebrook}, {McGregor}, {Damjanov},
  {Wisnioski}, {Abraham}, {Colless}, {Sharp}, {Crain}, {Poole}, \&
  {McCarthy}}]{green2014}
---. 2014,
  \href{http://dx.doi.org/10.1093/mnras/stt1882}{\JournalTitle{\mnras}, 437,
  1070}

\bibitem[{{Guo} {et~al.}(2012){Guo}, {Giavalisco}, {Ferguson}, {Cassata}, \&
  {Koekemoer}}]{guo2012}
{Guo}, Y., {Giavalisco}, M., {Ferguson}, H.~C., {Cassata}, P., \& {Koekemoer},
  A.~M. 2012,
  \href{http://dx.doi.org/10.1088/0004-637X/757/2/120}{\JournalTitle{\apj},
  757, 120}

\bibitem[{{Guo} {et~al.}(2015){Guo}, {Ferguson}, {Bell}, {Koo}, {Conselice},
  {Giavalisco}, {Kassin}, {Lu}, {Lucas}, {Mandelker}, {McIntosh}, {Primack},
  {Ravindranath}, {Barro}, {Ceverino}, {Dekel}, {Faber}, {Fang}, {Koekemoer},
  {Noeske}, {Rafelski}, \& {Straughn}}]{guo2015}
{Guo}, Y., {Ferguson}, H.~C., {Bell}, E.~F., {et~al.} 2015,
  \href{http://dx.doi.org/10.1088/0004-637X/800/1/39}{\JournalTitle{\apj}, 800,
  39}

\bibitem[{{Hao} {et~al.}(2011){Hao}, {Kennicutt}, {Johnson}, {Calzetti},
  {Dale}, \& {Moustakas}}]{hao2011}
{Hao}, C.-N., {Kennicutt}, R.~C., {Johnson}, B.~D., {et~al.}
  \href{http://dx.doi.org/10.1088/0004-637X/741/2/124}{2011, 741, 124}

\bibitem[{{Herrera-Camus} {et~al.}(2017){Herrera-Camus}, {Bolatto}, {Wolfire},
  {Ostriker}, {Draine}, {Leroy}, {Sandstrom}, {Hunt}, {Kennicutt}, {Calzetti},
  {Smith}, {Croxall}, {Galametz}, {de Looze}, {Dale}, {Crocker}, \&
  {Groves}}]{herreracamus2017}
{Herrera-Camus}, R., {Bolatto}, A., {Wolfire}, M., {et~al.} 2017,
  \href{http://dx.doi.org/10.3847/1538-4357/835/2/201}{\JournalTitle{\apj},
  835, 201}

\bibitem[{{Hodge} {et~al.}(2015){Hodge}, {Riechers}, {Decarli}, {Walter},
  {Carilli}, {Daddi}, \& {Dannerbauer}}]{hodge2015}
{Hodge}, J.~A., {Riechers}, D., {Decarli}, R., {et~al.} 2015,
  \href{http://dx.doi.org/10.1088/2041-8205/798/1/L18}{\JournalTitle{\apjl},
  798, L18}

\bibitem[{{Hopkins} \& {Beacom}(2006)}]{hopkinsbeacom2006}
{Hopkins}, A.~M., \& {Beacom}, J.~F.
  \href{http://dx.doi.org/10.1086/506610}{2006, 651, 142}

\bibitem[{{Hopkins} {et~al.}(2014){Hopkins}, {Kere{\v{s}}}, {O{\~n}orbe},
  {Faucher-Gigu{\`e}re}, {Quataert}, {Murray}, \& {Bullock}}]{hopkins2014}
{Hopkins}, P.~F., {Kere{\v{s}}}, D., {O{\~n}orbe}, J., {et~al.} 2014,
  \href{http://dx.doi.org/10.1093/mnras/stu1738}{\JournalTitle{\mnras}, 445,
  581}

\bibitem[{{Hopkins} {et~al.}(2011){Hopkins}, {Quataert}, \&
  {Murray}}]{hopkins2011model}
{Hopkins}, P.~F., {Quataert}, E., \& {Murray}, N.
  \href{http://dx.doi.org/10.1111/j.1365-2966.2011.19306.x}{2011, 417, 950}

\bibitem[{{Inoue} \& {Yoshida}(2018)}]{inoue2018}
{Inoue}, S., \& {Yoshida}, N. 2018,
  \href{http://dx.doi.org/10.1093/mnras/stx2978}{\JournalTitle{\mnras}, 474,
  3466}

\bibitem[{{Isobe} {et~al.}(1990){Isobe}, {Feigelson}, {Akritas}, \&
  {Babu}}]{isobe1990}
{Isobe}, T., {Feigelson}, E.~D., {Akritas}, M.~G., \& {Babu}, G.~J. 1990,
  \href{http://dx.doi.org/10.1086/169390}{\JournalTitle{\apj}, 364, 104}

\bibitem[{{Kennicutt}(1998)}]{kennicutt98}
{Kennicutt}, R.~C. \href{http://dx.doi.org/10.1086/305588}{1998, 498, 541}

\bibitem[{{Kim} {et~al.}(2011){Kim}, {Kim}, \& {Ostriker}}]{kim2011}
{Kim}, C.-G., {Kim}, W.-T., \& {Ostriker}, E.~C. 2011,
  \href{http://dx.doi.org/10.1088/0004-637X/743/1/25}{\JournalTitle{\apj}, 743,
  25}

\bibitem[{{Kim} {et~al.}(2013){Kim}, {Ostriker}, \& {Kim}}]{kim2013}
{Kim}, C.-G., {Ostriker}, E.~C., \& {Kim}, W.-T. 2013,
  \href{http://dx.doi.org/10.1088/0004-637X/776/1/1}{\JournalTitle{\apj}, 776,
  1}

\bibitem[{{Kim} {et~al.}(2017){Kim}, {Ostriker}, \& {Raileanu}}]{kim2017}
{Kim}, C.-G., {Ostriker}, E.~C., \& {Raileanu}, R. 2017,
  \href{http://dx.doi.org/10.3847/1538-4357/834/1/25}{\JournalTitle{\apj}, 834,
  25}

\bibitem[{{Kim} {et~al.}(2018){Kim}, {Kim}, \& {Ostriker}}]{kim2018rp}
{Kim}, J.-G., {Kim}, W.-T., \& {Ostriker}, E.~C. 2018,
  \href{http://dx.doi.org/10.3847/1538-4357/aabe27}{\JournalTitle{\apj}, 859,
  68}

\bibitem[{{Krumholz} \& {Burkhart}(2016)}]{krumholz2016}
{Krumholz}, M.~R., \& {Burkhart}, B. 2016,
  \href{http://dx.doi.org/10.1093/mnras/stw434}{\JournalTitle{mnras}, 458,
  1671}

\bibitem[{{Krumholz} {et~al.}(2017){Krumholz}, {Burkhart}, {Forbes}, \&
  {Crocker}}]{krumholz2017}
{Krumholz}, M.~R., {Burkhart}, B., {Forbes}, J.~C., \& {Crocker}, R.~M. 2017,
  \JournalTitle{ArXiv e-prints},
  \href{http://arxiv.org/abs/1706.00106}{{\sffamily arXiv:1706.00106}}

\bibitem[{{Krumholz} {et~al.}(2012){Krumholz}, {Dekel}, \&
  {McKee}}]{krumholz2012}
{Krumholz}, M.~R., {Dekel}, A., \& {McKee}, C.~F. 2012,
  \href{http://dx.doi.org/10.1088/0004-637X/745/1/69}{\JournalTitle{\apj}, 745,
  69}

\bibitem[{{Krumholz} \& {McKee}(2005)}]{krumholz2005}
{Krumholz}, M.~R., \& {McKee}, C.~F. 2005,
  \href{http://dx.doi.org/10.1086/431734}{\JournalTitle{\apj}, 630, 250}

\bibitem[{{Lagos} {et~al.}(2015){Lagos}, {Crain}, {Schaye}, {Furlong}, {Frenk},
  {Bower}, {Schaller}, {Theuns}, {Trayford}, {Bah{\'e}}, \& {Dalla
  Vecchia}}]{lagos2015}
{Lagos}, C.~d.~P., {Crain}, R.~A., {Schaye}, J., {et~al.} 2015,
  \href{http://dx.doi.org/10.1093/mnras/stv1488}{\JournalTitle{\mnras}, 452,
  3815}

\bibitem[{{Larkin} {et~al.}(2006){Larkin}, {Barczys}, {Krabbe}, {Adkins},
  {Aliado}, {Amico}, {Brims}, {Campbell}, {Canfield}, {Gasaway}, {Honey},
  {Iserlohe}, {Johnson}, {Kress}, {LaFreniere}, {Magnone}, {Magnone},
  {McElwain}, {Moon}, {Quirrenbach}, {Skulason}, {Song}, {Spencer}, {Weiss}, \&
  {Wright}}]{larkin2006}
{Larkin}, J., {Barczys}, M., {Krabbe}, A., {et~al.} 2006,
  \href{http://dx.doi.org/10.1016/j.newar.2006.02.005}{\JournalTitle{\nar}, 50,
  362}

\bibitem[{{Lehnert} {et~al.}(2013){Lehnert}, {Le Tiran}, {Nesvadba}, {van
  Driel}, {Boulanger}, \& {Di Matteo}}]{lehnert2013}
{Lehnert}, M.~D., {Le Tiran}, L., {Nesvadba}, N.~P.~H., {et~al.} 2013,
  \href{http://dx.doi.org/10.1051/0004-6361/201220555}{\JournalTitle{\aap},
  555, A72}

\bibitem[{{Lehnert} {et~al.}(2009){Lehnert}, {Nesvadba}, {Le Tiran}, {Di
  Matteo}, {van Driel}, {Douglas}, {Chemin}, \& {Bournaud}}]{lehnert2009}
{Lehnert}, M.~D., {Nesvadba}, N.~P.~H., {Le Tiran}, L., {et~al.} 2009,
  \href{http://dx.doi.org/10.1088/0004-637X/699/2/1660}{\JournalTitle{\apj},
  699, 1660}

\bibitem[{{Leroy} {et~al.}(2008){Leroy}, {Walter}, {Brinks}, {Bigiel}, {de
  Blok}, {Madore}, \& {Thornley}}]{leroy2008}
{Leroy}, A.~K., {Walter}, F., {Brinks}, E., {et~al.}
  \href{http://dx.doi.org/10.1088/0004-6256/136/6/2782}{2008, 136, 2782}

\bibitem[{{Leroy} {et~al.}(2012){Leroy}, {Bigiel}, {de Blok}, {Boissier},
  {Bolatto}, {Brinks}, {Madore}, {Munoz-Mateos}, {Murphy}, {Sandstrom},
  {Schruba}, \& {Walter}}]{leroy2012}
{Leroy}, A.~K., {Bigiel}, F., {de Blok}, W.~J.~G., {et~al.} 2012,
  \JournalTitle{ArXiv e-prints},
  \href{http://arxiv.org/abs/1202.2873}{{\sffamily arXiv:1202.2873
  [astro-ph.CO]}}

\bibitem[{{Levy} {et~al.}(2018){Levy}, {Bolatto}, {Teuben}, {S{\'a}nchez},
  {Barrera-Ballesteros}, {Blitz}, {Colombo}, {Garc{\'\i}a-Benito},
  {Herrera-Camus}, {Husemann}, {Kalinova}, {Lan}, {Leung}, {Mast}, {Utomo},
  {van de Ven}, {Vogel}, \& {Wong}}]{levy2018}
{Levy}, R.~C., {Bolatto}, A.~D., {Teuben}, P., {et~al.} 2018,
  \href{http://dx.doi.org/10.3847/1538-4357/aac2e5}{\JournalTitle{\apj}, 860,
  92}

\bibitem[{{Madau} \& {Dickinson}(2014)}]{madau2014}
{Madau}, P., \& {Dickinson}, M. 2014,
  \href{http://dx.doi.org/10.1146/annurev-astro-081811-125615}{\JournalTitle{\araa},
  52, 415}

\bibitem[{{Magdis} {et~al.}(2017){Magdis}, {Rigopoulou}, {Daddi}, {Bethermin},
  {Feruglio}, {Sargent}, {Dannerbauer}, {Dickinson}, {Elbaz}, {Gomez Guijarro},
  {Huang}, {Toft}, \& {Valentino}}]{magdis2017}
{Magdis}, G.~E., {Rigopoulou}, D., {Daddi}, E., {et~al.} 2017,
  \href{http://dx.doi.org/10.1051/0004-6361/201731037}{\JournalTitle{\aap},
  603, A93}

\bibitem[{{Moiseev} {et~al.}(2015){Moiseev}, {Tikhonov}, \&
  {Klypin}}]{Moiseev2015}
{Moiseev}, A.~V., {Tikhonov}, A.~V., \& {Klypin}, A. 2015,
  \href{http://dx.doi.org/10.1093/mnras/stv489}{\JournalTitle{\mnras}, 449,
  3568}

\bibitem[{{Murray} {et~al.}(2011){Murray}, {M{\'e}nard}, \&
  {Thompson}}]{murray2011}
{Murray}, N., {M{\'e}nard}, B., \& {Thompson}, T.~A. 2011,
  \href{http://dx.doi.org/10.1088/0004-637X/735/1/66}{\JournalTitle{\apj}, 735,
  66}

\bibitem[{{Nanayakkara} {et~al.}(2017){Nanayakkara}, {Glazebrook}, {Kacprzak},
  {Yuan}, {Fisher}, {Tran}, {Kewley}, {Spitler}, {Alcorn}, {Cowley}, {Labbe},
  {Straatman}, \& {Tomczak}}]{nanayakkara2017}
{Nanayakkara}, T., {Glazebrook}, K., {Kacprzak}, G.~G., {et~al.} 2017,
  \href{http://dx.doi.org/10.1093/mnras/stx605}{\JournalTitle{\mnras}, 468,
  3071}

\bibitem[{{Narayanan} {et~al.}(2011){Narayanan}, {Krumholz}, {Ostriker}, \&
  {Hernquist}}]{narayanan2011}
{Narayanan}, D., {Krumholz}, M., {Ostriker}, E.~C., \& {Hernquist}, L.
  \href{http://dx.doi.org/10.1111/j.1365-2966.2011.19516.x}{2011, 418, 664}

\bibitem[{{Narayanan} {et~al.}(2012){Narayanan}, {Krumholz}, {Ostriker}, \&
  {Hernquist}}]{narayanan2012}
{Narayanan}, D., {Krumholz}, M.~R., {Ostriker}, E.~C., \& {Hernquist}, L. 2012,
  \href{http://dx.doi.org/10.1111/j.1365-2966.2012.20536.x}{\JournalTitle{\mnras},
  421, 3127}

\bibitem[{{Obreschkow} \& {Rawlings}(2009)}]{obreschkow2009}
{Obreschkow}, D., \& {Rawlings}, S. 2009,
  \href{http://dx.doi.org/10.1088/0004-637X/696/2/L129}{\JournalTitle{\apjl},
  696, L129}

\bibitem[{{Obreschkow} {et~al.}(2015){Obreschkow}, {Glazebrook}, {Bassett},
  {Fisher}, {Abraham}, {Wisnioski}, {Green}, {McGregor}, {Damjanov}, {Popping},
  \& {Jorgensen}}]{obreschkow2015}
{Obreschkow}, D., {Glazebrook}, K., {Bassett}, R., {et~al.} 2015,
  \JournalTitle{ArXiv e-prints},
  \href{http://arxiv.org/abs/1508.04768}{{\sffamily arXiv:1508.04768}}

\bibitem[{{Oliva-Altamirano} {et~al.}(2017){Oliva-Altamirano}, {Fisher},
  {Glazebrook}, {Wisnioski}, {Bekiaris}, {Bassett}, {Obreschkow}, \&
  {Abraham}}]{oliva2017}
{Oliva-Altamirano}, P., {Fisher}, D., {Glazebrook}, K., {et~al.} 2017,
  \JournalTitle{ArXiv e-prints},
  \href{http://arxiv.org/abs/1710.09457}{{\sffamily arXiv:1710.09457}}

\bibitem[{{Ostriker} {et~al.}(2010){Ostriker}, {McKee}, \&
  {Leroy}}]{ostriker2010}
{Ostriker}, E.~C., {McKee}, C.~F., \& {Leroy}, A.~K.
  \href{http://dx.doi.org/10.1088/0004-637X/721/2/975}{2010, 721, 975}

\bibitem[{{Ostriker} \& {Shetty}(2011)}]{ostriker2011}
{Ostriker}, E.~C., \& {Shetty}, R. 2011,
  \href{http://dx.doi.org/10.1088/0004-637X/731/1/41}{\JournalTitle{\apj}, 731,
  41}

\bibitem[{{Rahman} {et~al.}(2012){Rahman}, {Bolatto}, {Xue}, {Wong}, {Leroy},
  {Walter}, {Bigiel}, {Rosolowsky}, {Fisher}, {Vogel}, {Blitz}, {West}, \&
  {Ott}}]{rahman2012}
{Rahman}, N., {Bolatto}, A.~D., {Xue}, R., {et~al.}
  \href{http://dx.doi.org/10.1088/0004-637X/745/2/183}{2012, 745, 183}

\bibitem[{{Robotham} \& {Obreschkow}(2015)}]{robotham2015}
{Robotham}, A.~S.~G., \& {Obreschkow}, D. 2015,
  \href{http://dx.doi.org/10.1017/pasa.2015.33}{\JournalTitle{\pasa}, 32, e033}

\bibitem[{{Saintonge} {et~al.}(2011){Saintonge}, {Kauffmann}, {Wang}, {Kramer},
  {Tacconi}, {Buchbender}, {Catinella}, {Graci{\'a}-Carpio}, {Cortese},
  {Fabello}, {Fu}, {Genzel}, {Giovanelli}, {Guo}, {Haynes}, {Heckman},
  {Krumholz}, {Lemonias}, {Li}, {Moran}, {Rodriguez-Fernandez}, {Schiminovich},
  {Schuster}, \& {Sievers}}]{saintonge2011b}
{Saintonge}, A., {Kauffmann}, G., {Wang}, J., {et~al.} 2011,
  \href{http://dx.doi.org/10.1111/j.1365-2966.2011.18823.x}{\JournalTitle{\mnras},
  415, 61}

\bibitem[{{Salim} {et~al.}(2015){Salim}, {Federrath}, \& {Kewley}}]{salim2015}
{Salim}, D.~M., {Federrath}, C., \& {Kewley}, L.~J. 2015,
  \href{http://dx.doi.org/10.1088/2041-8205/806/2/L36}{\JournalTitle{\apjl},
  806, L36}

\bibitem[{{Sandstrom} {et~al.}(2013){Sandstrom}, {Leroy}, {Walter}, {Bolatto},
  {Croxall}, {Draine}, {Wilson}, {Wolfire}, {Calzetti}, {Kennicutt}, {Aniano},
  {Donovan Meyer}, {Usero}, {Bigiel}, {Brinks}, {de Blok}, {Crocker}, {Dale},
  {Engelbracht}, {Galametz}, {Groves}, {Hunt}, {Koda}, {Kreckel}, {Linz},
  {Meidt}, {Pellegrini}, {Rix}, {Roussel}, {Schinnerer}, {Schruba}, {Schuster},
  {Skibba}, {van der Laan}, {Appleton}, {Armus}, {Brandl}, {Gordon}, {Hinz},
  {Krause}, {Montiel}, {Sauvage}, {Schmiedeke}, {Smith}, \&
  {Vigroux}}]{sandstrom2013}
{Sandstrom}, K.~M., {Leroy}, A.~K., {Walter}, F., {et~al.}
  \href{http://dx.doi.org/10.1088/0004-637X/777/1/5}{2013, 777, 5}

\bibitem[{{Schinnerer} {et~al.}(2016){Schinnerer}, {Groves}, {Sargent},
  {Karim}, {Oesch}, {Magnelli}, {LeFevre}, {Tasca}, {Civano}, {Cassata}, \&
  {Smol{\v c}i{\'c}}}]{schinnerer2016}
{Schinnerer}, E., {Groves}, B., {Sargent}, M.~T., {et~al.} 2016,
  \href{http://dx.doi.org/10.3847/1538-4357/833/1/112}{\JournalTitle{\apj},
  833, 112}

\bibitem[{{Scoville} {et~al.}(2016){Scoville}, {Sheth}, {Aussel}, {Vanden
  Bout}, {Capak}, {Bongiorno}, {Casey}, {Murchikova}, {Koda},
  {{\'A}lvarez-M{\'a}rquez}, {Lee}, {Laigle}, {McCracken}, {Ilbert}, {Pope},
  {Sanders}, {Chu}, {Toft}, {Ivison}, \& {Manohar}}]{scoville2016}
{Scoville}, N., {Sheth}, K., {Aussel}, H., {et~al.} 2016,
  \href{http://dx.doi.org/10.3847/0004-637X/820/2/83}{\JournalTitle{\apj}, 820,
  83}

\bibitem[{{Shetty} \& {Ostriker}(2012)}]{shetty2012}
{Shetty}, R., \& {Ostriker}, E.~C. 2012,
  \href{http://dx.doi.org/10.1088/0004-637X/754/1/2}{\JournalTitle{\apj}, 754,
  2}

\bibitem[{{Speagle} {et~al.}(2014){Speagle}, {Steinhardt}, {Capak}, \&
  {Silverman}}]{speagle2014}
{Speagle}, J.~S., {Steinhardt}, C.~L., {Capak}, P.~L., \& {Silverman}, J.~D.
  2014,
  \href{http://dx.doi.org/10.1088/0067-0049/214/2/15}{\JournalTitle{\apjs},
  214, 15}

\bibitem[{{Spilker} {et~al.}(2016){Spilker}, {Marrone}, {Aravena},
  {B{\'e}thermin}, {Bothwell}, {Carlstrom}, {Chapman}, {Crawford}, {de Breuck},
  {Fassnacht}, {Gonzalez}, {Greve}, {Hezaveh}, {Litke}, {Ma}, {Malkan},
  {Rotermund}, {Strandet}, {Vieira}, {Weiss}, \& {Welikala}}]{spilker2016}
{Spilker}, J.~S., {Marrone}, D.~P., {Aravena}, M., {et~al.} 2016,
  \href{http://dx.doi.org/10.3847/0004-637X/826/2/112}{\JournalTitle{\apj},
  826, 112}

\bibitem[{{Swinbank} {et~al.}(2012){Swinbank}, {Smail}, {Sobral}, {Theuns},
  {Best}, \& {Geach}}]{swinbank2012}
{Swinbank}, A.~M., {Smail}, I., {Sobral}, D., {et~al.} 2012,
  \href{http://dx.doi.org/10.1088/0004-637X/760/2/130}{\JournalTitle{\apj},
  760, 130}

\bibitem[{{Swinbank} {et~al.}(2011){Swinbank}, {Papadopoulos}, {Cox}, {Krips},
  {Ivison}, {Smail}, {Thomson}, {Neri}, {Richard}, \& {Ebeling}}]{swinbank2011}
{Swinbank}, A.~M., {Papadopoulos}, P.~P., {Cox}, P., {et~al.} 2011,
  \href{http://dx.doi.org/10.1088/0004-637X/742/1/11}{\JournalTitle{\apj}, 742,
  11}

\bibitem[{{Swinbank} {et~al.}(2017){Swinbank}, {Harrison}, {Trayford},
  {Schaller}, {Smail}, {Schaye}, {Theuns}, {Smit}, {Alexander}, {Bacon},
  {Bower}, {Contini}, {Crain}, {de Breuck}, {Decarli}, {Epinat}, {Fumagalli},
  {Furlong}, {Galametz}, {Johnson}, {Lagos}, {Richard}, {Vernet}, {Sharples},
  {Sobral}, \& {Stott}}]{swinbank2017}
{Swinbank}, A.~M., {Harrison}, C.~M., {Trayford}, J., {et~al.} 2017,
  \href{http://dx.doi.org/10.1093/mnras/stx201}{\JournalTitle{\mnras}, 467,
  3140}

\bibitem[{{Tacconi} {et~al.}(2013){Tacconi}, {Neri}, {Genzel}, {Combes},
  {Bolatto}, {Cooper}, {Wuyts}, {Bournaud}, {Burkert}, {Comerford}, {Cox},
  {Davis}, {F{\"o}rster Schreiber}, {Garc{\'{\i}}a-Burillo}, {Gracia-Carpio},
  {Lutz}, {Naab}, {Newman}, {Omont}, {Saintonge}, {Shapiro Griffin}, {Shapley},
  {Sternberg}, \& {Weiner}}]{tacconi2013}
{Tacconi}, L.~J., {Neri}, R., {Genzel}, R., {et~al.} 2013,
  \href{http://dx.doi.org/10.1088/0004-637X/768/1/74}{\JournalTitle{\apj}, 768,
  74}

\bibitem[{{Tacconi} {et~al.}(2017){Tacconi}, {Genzel}, {Saintonge}, {Combes},
  {Garc{\'{\i}}a-Burillo}, {Neri}, {Bolatto}, {Contini}, {F{\"o}rster
  Schreiber}, {Lilly}, {Lutz}, {Wuyts}, {Accurso}, {Boissier}, {Boone},
  {Bouch{\'e}}, {Bournaud}, {Burkert}, {Carollo}, {Cooper}, {Cox}, {Feruglio},
  {Freundlich}, {Herrera-Camus}, {Juneau}, {Lippa}, {Naab}, {Renzini},
  {Salome}, {Sternberg}, {Tadaki}, {{\"U}bler}, {Walter}, {Weiner}, \&
  {Weiss}}]{tacconi2017}
{Tacconi}, L.~J., {Genzel}, R., {Saintonge}, A., {et~al.} 2017,
  \JournalTitle{ArXiv e-prints},
  \href{http://arxiv.org/abs/1702.01140}{{\sffamily arXiv:1702.01140}}

\bibitem[{{Toomre}(1964)}]{toomre1964}
{Toomre}, A. 1964, \href{http://dx.doi.org/10.1086/147861}{\JournalTitle{\apj},
  139, 1217}

\bibitem[{{{\"U}bler} {et~al.}(2018){{\"U}bler}, {Genzel}, {Tacconi},
  {F{\"o}rster Schreiber}, {Neri}, {Contursi}, {Belli}, {Nelson}, {Lang},
  {Shimizu}, {Davies}, {Herrera-Camus}, {Lutz}, {Plewa}, {Price}, {Schuster},
  {Sternberg}, {Tadaki}, {Wisnioski}, \& {Wuyts}}]{ubler2018}
{{\"U}bler}, H., {Genzel}, R., {Tacconi}, L.~J., {et~al.} 2018,
  \href{http://dx.doi.org/10.3847/2041-8213/aaacfa}{\JournalTitle{\apj}, 854,
  L24}

\bibitem[{{Walter} {et~al.}(2008){Walter}, {Brinks}, {de Blok}, {Bigiel},
  {Kennicutt}, {Thornley}, \& {Leroy}}]{walter2008}
{Walter}, F., {Brinks}, E., {de Blok}, W.~J.~G., {et~al.} 2008,
  \href{http://dx.doi.org/10.1088/0004-6256/136/6/2563}{\JournalTitle{\aj},
  136, 2563}

\bibitem[{{White} {et~al.}(2017){White}, {Fisher}, {Murray}, {Glazebrook},
  {Abraham}, {Bolatto}, {Green}, {Mentuch Cooper}, \& {Obreschkow}}]{white2017}
{White}, H.~A., {Fisher}, D.~B., {Murray}, N., {et~al.} 2017,
  \href{http://dx.doi.org/10.3847/1538-4357/aa7fbf}{\JournalTitle{\apj}, 846,
  35}

\bibitem[{{Wisnioski} {et~al.}(2012){Wisnioski}, {Glazebrook}, {Blake},
  {Poole}, {Green}, {Wyder}, \& {Martin}}]{wisnioski2012}
{Wisnioski}, E., {Glazebrook}, K., {Blake}, C., {et~al.} 2012,
  \href{http://dx.doi.org/10.1111/j.1365-2966.2012.20850.x}{\JournalTitle{\mnras},
  422, 3339}

\bibitem[{{Wisnioski} {et~al.}(2015){Wisnioski}, {F{\"o}rster Schreiber},
  {Wuyts}, {Wuyts}, {Bandara}, {Wilman}, {Genzel}, {Bender}, {Davies},
  {Fossati}, {Lang}, {Mendel}, {Beifiori}, {Brammer}, {Chan}, {Fabricius},
  {Fudamoto}, {Kulkarni}, {Kurk}, {Lutz}, {Nelson}, {Momcheva}, {Rosario},
  {Saglia}, {Seitz}, {Tacconi}, \& {van Dokkum}}]{wisnioski2015}
{Wisnioski}, E., {F{\"o}rster Schreiber}, N.~M., {Wuyts}, S., {et~al.} 2015,
  \href{http://dx.doi.org/10.1088/0004-637X/799/2/209}{\JournalTitle{\apj},
  799, 209}

\end{thebibliography}

\newpage
\appendix

\section{CO Spectra and Observational Details }
In Table~3 we list the observation parameters and derived fluxes for CO(1-0) observations of DYNAMO galaxies. The observations were carried out in three separate campaigns with the Plateu de Bur\'e Interferometer, also called NOEMA. All observations were made with the WIDEX system. Each observing campaign had similar sensitivity goals of $\sim$1.5~mJy in 50~km~s$^{-1}$ channels.  Of the CO measurements used in this paper 8 have been published in previous work \citep{fisher2014,white2017}. Details of those corresponding observations are also outlined in those papers. 

\begin{figure}
\begin{center}
\includegraphics[width=0.3\textwidth]{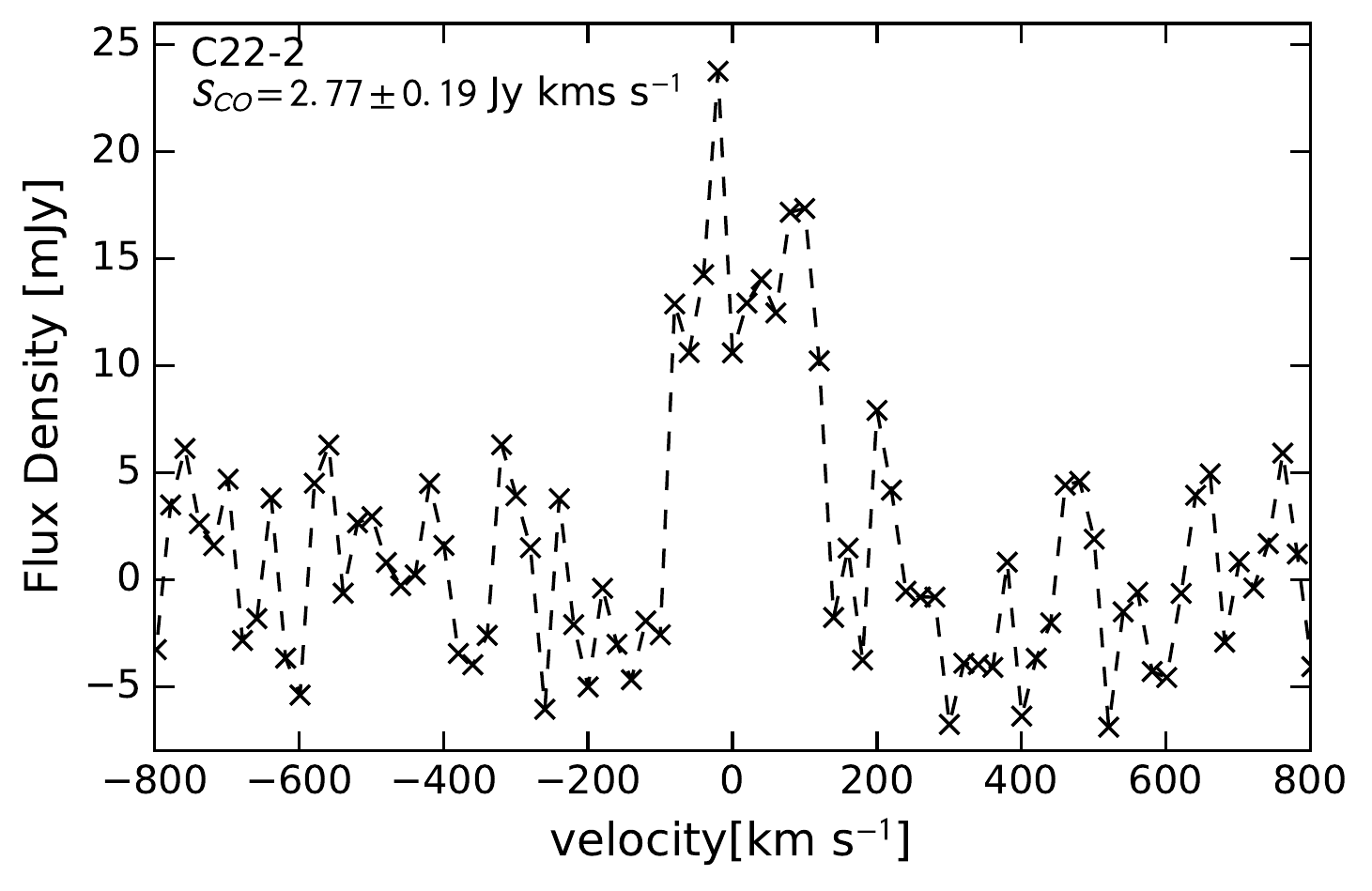}
\includegraphics[width=0.3\textwidth]{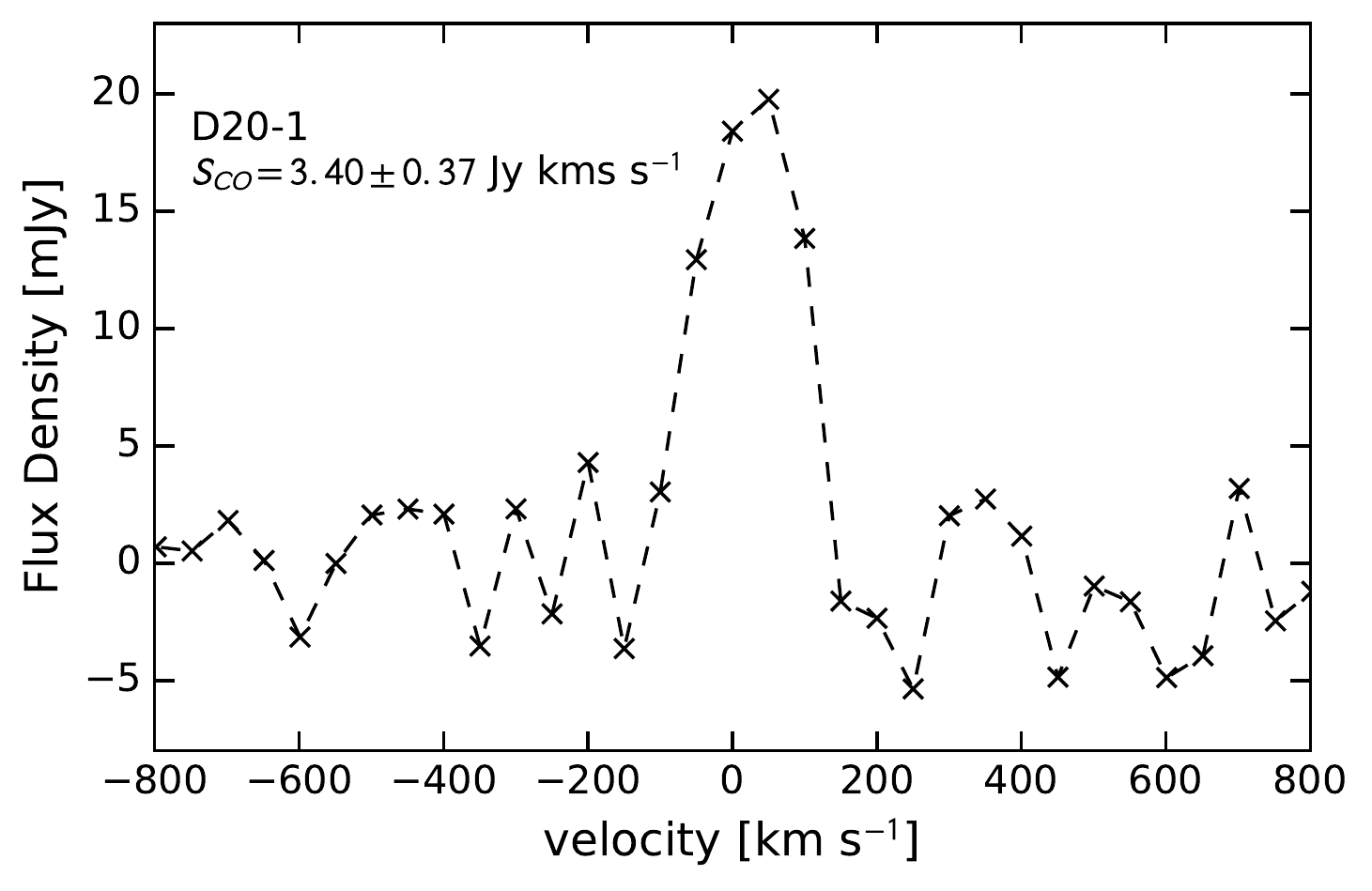}
\includegraphics[width=0.3\textwidth]{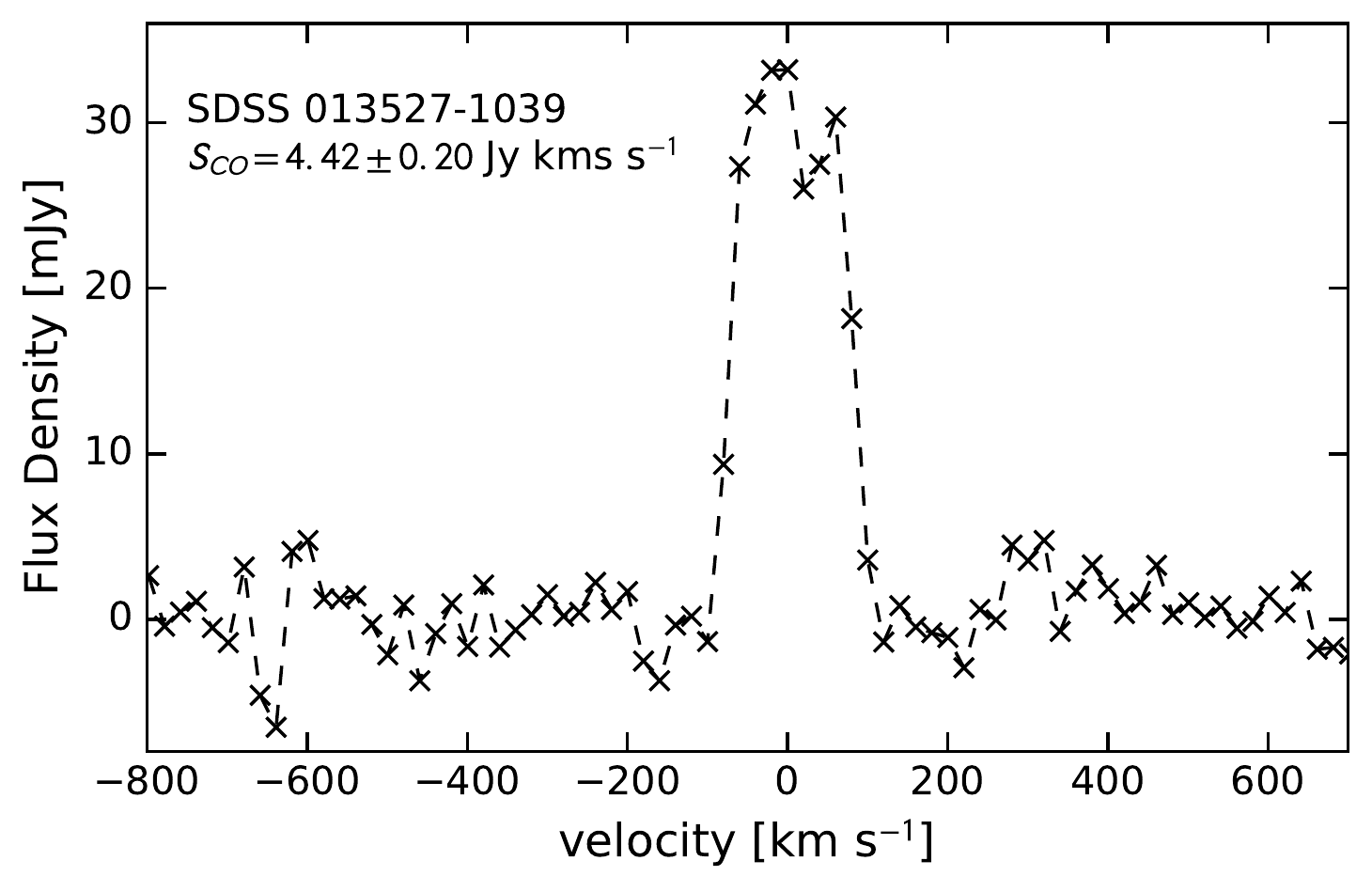}
\includegraphics[width=0.3\textwidth]{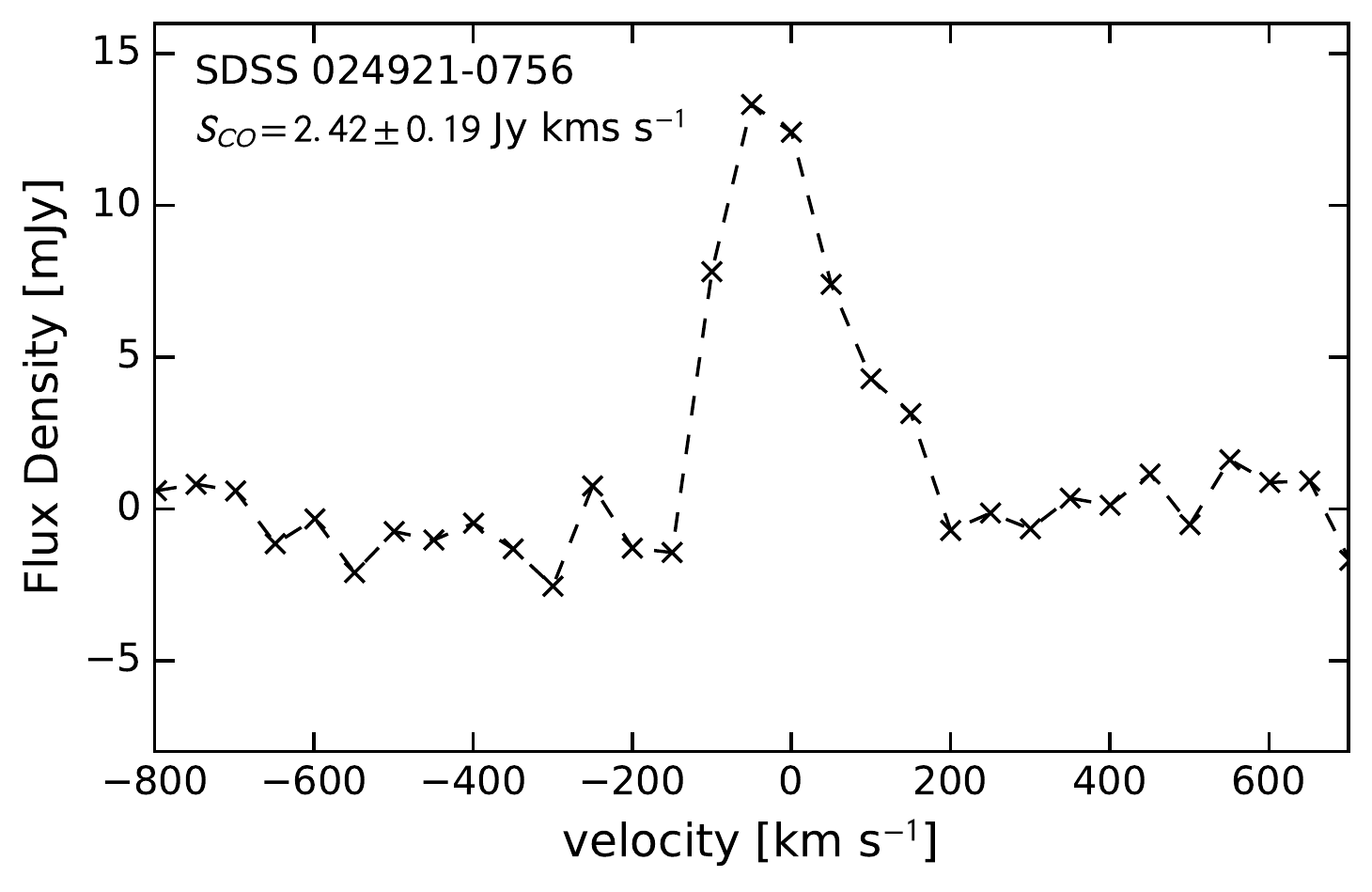}
\includegraphics[width=0.3\textwidth]{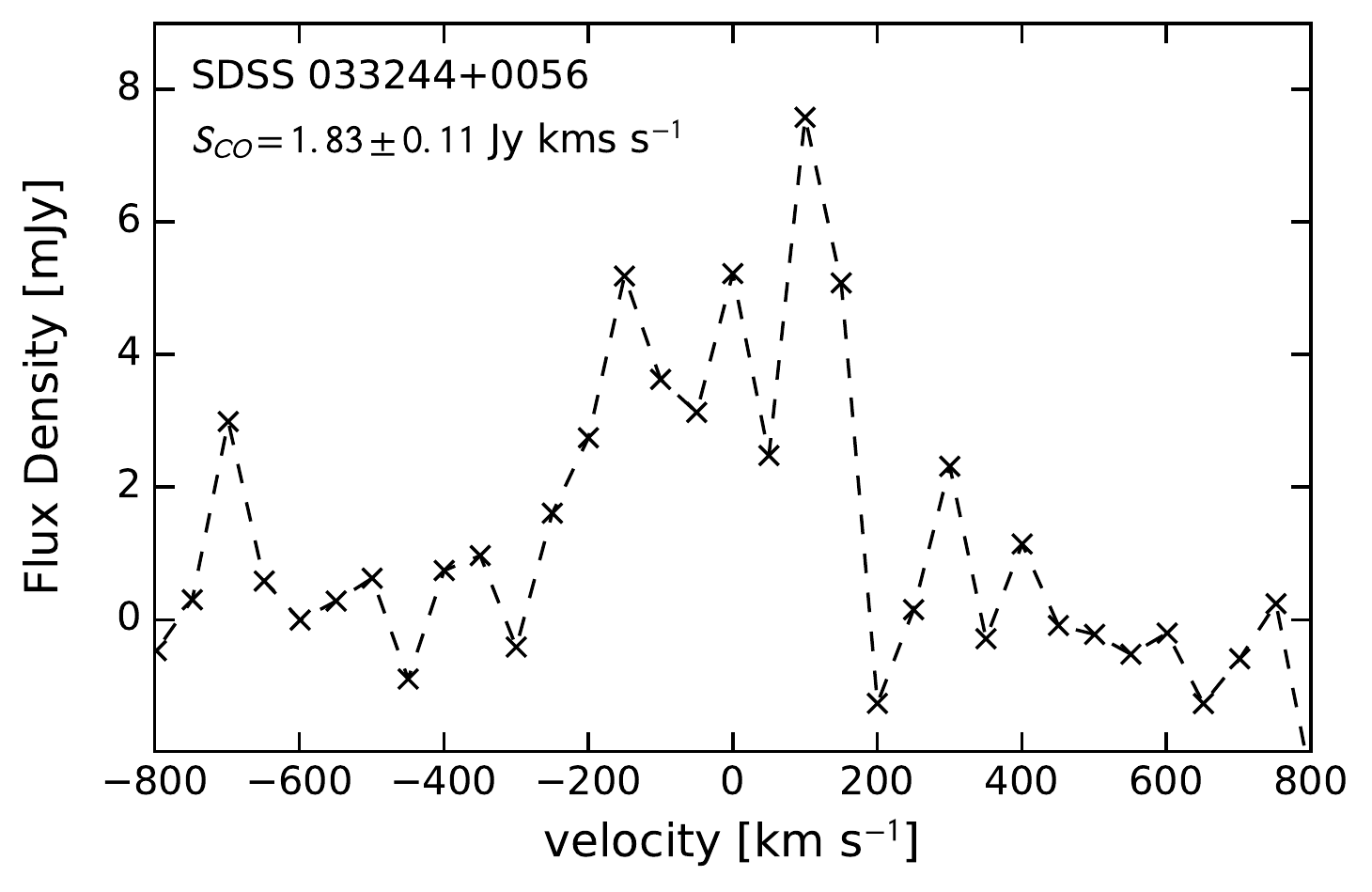}
\includegraphics[width=0.3\textwidth]{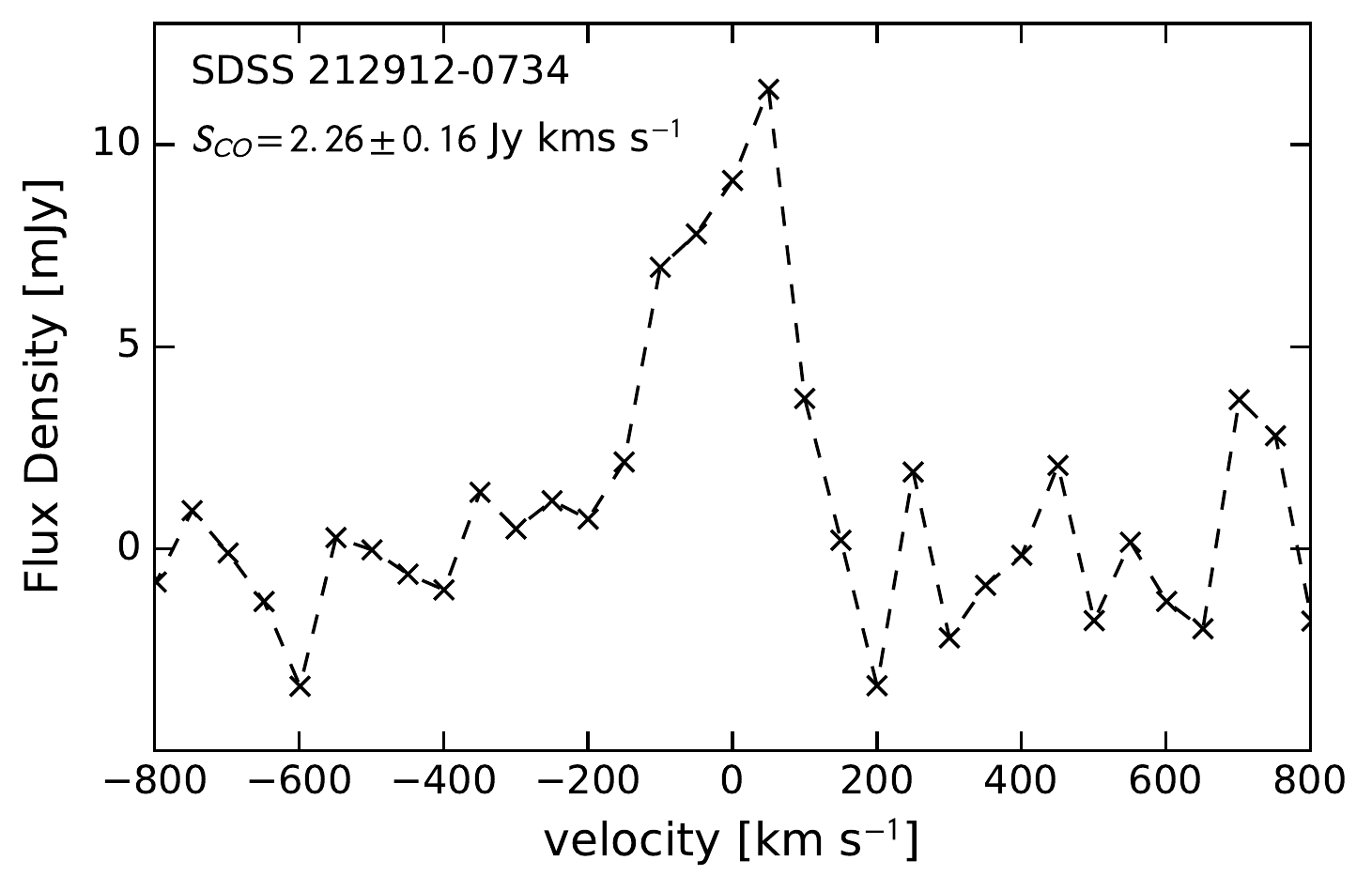}
\end{center}
\caption{The above spectra represent data for new NOEMA observations used in this work. Each spectrum is centered on the red-shifted CO(1-0) emission line.  All flux densities are measured using a binning of 50~km~s$^{-1}$. \label{fig:cospect}}
\end{figure}

Spectra of new observations are shown in Fig.~\ref{fig:cospect}. Similar spectra for previously published observations are given in the respective publications. For each new observations we plot the flux density in mJy against the velocity in km~s$^{-1}$. The redshifted CO(1-0) transition is centered at the velocity of 0~km~s$^{-1}$. All fluxes are determined by binning the spectra into 50~km~s$^{-1}$ channels, as described in the methods section.

\begin{deluxetable}{lccccccc}
\tablewidth{0pt} \tablecaption{CO Observations }
\tablehead{ \colhead{Galaxy} & \colhead{Observation } & \colhead{Time on Target}  & \colhead{Beam Size} & \colhead{$\nu_{CO}$(sky)} & \colhead{Line width} & \colhead{$F_{CO}$}& \colhead{Source}  \\
\colhead{ } & \colhead{ Date} & \colhead{[hr]} & \colhead{arcsec$^2$} &\colhead{ [GHz] } & \colhead{ [km s$^{-1}$] } & \colhead{[Jy km s$^{-1}$] } & \colhead{Source}   
}
\startdata
G10-1 & 11-Jun-13 & 1.0 & 5.86 $\times$ 4.55 & 100.786 & 358 & 1.6 $\pm$ 0.26 & \cite{fisher2014} \\
 & 17-Jun-13 & 1.4 &    &  &  &    &  \\
D13-5 & 30-May-13 & 1.8 & 6.32 $\times$ 3.54 & 107.194 & 334 & 10.04 $\pm$ 0.31 & \cite{fisher2014} \\
G04-1 & 21-Jun-13 & 0.8 & 10.94 $\times$ 5.25 & 102.027 & 352 & 6.63 $\pm$ 0.48 & \cite{fisher2014} \\
 & 16-Jul-13 & 0.8 &    &  &  &    &  \\
G20-2 & 23-May-14 & 0.9 & 9.46 $\times$ 4.71 & 101.017 & 237 & 1.57 $\pm$ 0.18 & \cite{white2017} \\
G08-5 & 20-May-14 & 1.1 & 5.36 $\times$ 4.47 & 101.812 & 353 & 2.44 $\pm$ 0.27 & \cite{white2017} \\
D15-3 & 30-May-14  & 1.5 & 6.26 $\times$ 4.44 & 108.023 & 361 & 12.8 $\pm$ 0.25 & \cite{white2017} \\
G14-1 & 29-May-14  & 1.1 & 7.13 $\times$ 4.66 & 101.803 & 236 & 1.69 $\pm$ 0.2 & \cite{white2017} \\
C13-1 & 20-May-14 & 1.1 & 5.94 $\times$ 4.19 & 106.851 & 196 & 5.84 $\pm$ 0.15 & \cite{white2017} \\
C22-2 & 19-Jul-16 & 1.9 & 34.87 $\times$ 2.73 & 107.613 & 240 & 2.77 $\pm$ 0.19 & This Work \\
D20-1 & 19-Jul-16 & 1.1 & 56.5 $\times$ 6.31 & 107.681 & 280 & 3.4 $\pm$ 0.37 & This Work \\
SDSS024921-0756 & 15-Aug-16 & 2.2 & 5.3 $\times$ 3.16 & 99.975 & 300 & 2.42 $\pm$ 0.19 & This Work \\
SDSS212912-0734 & 9-Jul-16 & 1.5 & 4.91 $\times$ 3.89 & 97.357 & 340 & 2.26 $\pm$ 0.16 & This Work \\
 & 10-Jul-16 & 3.8 &    &  &  &    &  \\
SDSS013527-1039 & 02-Dec-16 & 3.4 & 4.57 $\times$ 1.689 & 102.281 & 220 & 4.42 $\pm$ 0.2 & This Work \\
SDSS033244+0056 & 10-Dec-16  & 1.9 & 3.39 $\times$ 2.17 & 97.522 & 460 & 1.83 $\pm$ 0.11 & This Work \\
\enddata

\end{deluxetable}

\section{Estimation of Physical Parameters for Measurement of Pressure }

\subsection{Galaxy Sizes}
The size of the star light is measured from the continuum observations corresponding to each emission line measurement. Similarly the sizes used to calculate the SFR surface density is measured from the resolved emission line maps for each target. For THINGS galaxies \cite{leroy2008} measures the sizes in stars, SFR and molecular gas, and for those targets we use the corresponding measurement to directly measure the associated surface brightness. Our observations of CO(1-0) in most DYNAMO targets are unresolved source detections, with a handful that are marginally resolved, in which $R_{disk}$ is only slightly larger than the beam size. To calculate the total mid-plane pressure for DYNAMO galaxies we, therefore, must make an assumption on the sizes of the molecular gas. 

A straightforward method to estimate the molecular gas disk is to assume that the half-light radius of the ionized gas is roughly equivalent as that of the molecular gas. In the local Universe the surface brightness profiles of gas in disk galaxies have been shown to be well behaved with a regular, exponential decay \citep{bigiel2012}. In the THINGS sample \cite{leroy2008} finds that the average ratio of CO-to-SFR scale-lengths is $<l_{CO}/l_{SFR}=1.0\pm0.2$. Assuming that the distribution of ionized gas is a good proxy for the distribution of the star formation, the assumption that $R_{1/2}(CO)\approx R_{1/2}(H\alpha)$ would be well justified in similar galaxies. 

Due to the difficulties of such observations there is considerably less work comparing the distribution of molecular gas to the distribution of stars, or star formation in turbulent disks. \cite{hodge2015} measure the the size of gas, CO(2-1), and star formation, 880~$\mu$m, in a rare double-lensed system at $z\sim4$ with $\sim 1$~kpc resolution, finding that the respective physical size of the disk is 14~kpc in CO and $\sim$10~kpc in star formation. \cite{bolatto2015} studies high-resolution maps of CO(1-0) and CO(3-2) in two $z\sim2$ targets, finding that the CO gas in these targets $R_{1/2}(CO)-R_{1/2}(optical)\approx$1~kpc. Similarly, the sample of \cite{tacconi2013}, which has considerably lower spatial resolution, nonetheless shows that on average $R_{1/2}(CO)/R_{1/2}(optical)\approx 1$ with standard deviation of 0.5. \cite{dessauges2015} map CO(2-1) in a sample of strongly lensed galaxies at $z\sim2$ and find $R_{1/2}(CO)\sim 1-4$~kpc, which is similar to our estimates from H$\alpha$. We note that more extreme differences have been observed, however those targets are typically found to have multiple velocity components indicating they are likely ongoing mergers \citep[e.g.][]{spilker2016}. Overall, we the observations of main-sequence galaxies at $z>1$ seem to suggest that the uncertainty on our estimation of $R_{1/2}(CO)\approx R_{1/2}(H\alpha)$ would be at the $\sim$20-50\% level. Increasing the size of the molecular gas disk by 50\% (i.e. $R=1.5\ R_{1/2}(H\alpha)$) would decrease our measured pressures by a factor of $2-3\times$ in DYNAMO galaxies. 

For our unresolved observations we can use the measured beam size as an estimate of the ``largest possible size" for the molecular gas disk in DYNAMO galaxies, which will then correspond to a lowest possible pressure. For the 5 galaxies in which the beam is slightly smaller than $R_{disk}$ we use the radius at which the flux from the galaxy is equivalent to the noise. The median for DYNAMO galaxies the circularized largest possible CO disk radius is $\sim 1.6$ times larger than the corresponding $R_{disk}$ measured from the ionized gas. This corresponds to a decrease in the measured mid-plane pressure by a factor of 3-4.  

The THINGS observations suggest that for normal spirals, low pressure disks, assuming the molecular gas and star formation have similar disk sizes is a safe approximations. Observations of higher redshift sources suggest this is roughly a safe assumption, though molecular disks in these higher pressure systems may be slightly larger. We opt for the simple assumption that $R_{disk}(CO)\approx R_{disk}(ion)$, as this is consistent with the data and allows us to make a single correction for low and high pressure systems. We then use the maximum observed CO(1-0) size for each target as the lower bound error on pressure that is introduced from the CO observations. This uncertainty will be added in quadrature with other uncertainties to determine the lower-limit of pressure for each target.

\subsection{Stellar Velocity Dispersion}
Based on results from \cite{bassett2014}, which compares the velocity dispersion of gas and stars in DYNAMO galaxies, we find that the standard approximation for stellar velocity dispersion, $\sigma_* \approx 1/2 \sqrt{\pi G l_{*} \Sigma_*}$, where the disk scale-length $l_{*}\approx R_{half,*}/1.76$, reproduced measured velocities within $\pm 10$~km~s$^{-1}$.  We also consider a simpler formulation where $\sigma_* \approx \sigma+ 15$~km~s$^{-1}$. We find that these result in similar overall values of $P$ when inserted into Equation~\ref{eq:mpp}. For the sake of consistency with previous studies \citep[e.g.][]{leroy2008} we use $\sigma_* \approx 1/2 \sqrt{\pi G l_{*} \Sigma_*}$. Note as done in \cite{bassett2014} we assume that for DYNAMO galaxies $\sigma\approx\sigma_{z}$. More work on the stellar kinematics of turbulent disks would certainly be informative for these assumptions. 

\subsection{Total Gas Surface Density}
A significant source uncertainty in the calculation of the mid-plane pressure is the estimation of the total gas mass surface density. Our observations of CO(1-0) only allow for observation of the molecular gas. However, estimations of mid-plane pressure refer to the entire gas mass surface density, atomic and molecular. Observing atomic gas masses on $z\sim0.1$ galaxies is difficult with present facilities \citep{catinella2015}, and is a primary goal of future SKA pathfinders. We caution that since the mid-plane pressure depends on $\Sigma_{gas}^2$, even small differences in $\Sigma_{gas}$ may significantly affect the slope of correlations.

We use a multi-step method to estimate the total gas density. Observations of local spirals find that $\Sigma_{atm}\sim 5-10$~M$_{\odot}$~pc$^{-2}$ \citep[e.g][]{bigiel2008}. We first use a conservative estimate assuming the constant $\Sigma_{atm}\sim 5$~M$_{\odot}$~pc$^{-2}$ to estimate the mid-plane pressure. \cite{blitzrosolowsky2006} give a correlation between the ratio of molecular-to-atomic gas, $R_{mol}$, and the total pressure, $P/k$. Using our initial estimate of pressure, which assumed, $\Sigma_{atm}\sim 5$~M$_{\odot}$~pc$^{-2}$, we then calculate the expected $R_{mol}$ for DYNAMO galaxies. Note that in DYNAMO galaxies the stellar and molecular surface densities are more likely to drive the value of the pressure.  We then re-calculate the mid-plane pressure with the new estimate of $\Sigma_{atm}= \Sigma_{mol}/R_{mol}$.  We find that for galaxies with high surface densities of gas ($\Sigma_{mol}\gtrsim 30$~M$_{\odot}$~pc$^{-2}$) the difference in calculated mid-plane pressure (constant versus variable $\Sigma_{atm}$) is small, less than 0.01 dex. However for the two lowest surface density galaxies in our sample the difference in pressures reach 0.12 dex. This difference will be reflected in error bars in the associated figures. We rerun this estimation assuming the larger value of $\Sigma_{atm}= 15$~M$_{\odot}$~pc$^{-2}$ as an initial guess. We find this has at most a difference of 0.04~dex in determined pressure, and only on the two targets with the lowest $\Sigma_{gas}$.

\end{document}